\documentclass[journal,draft,10 pt, a4paper,onecolumn]{IEEEtran}

\newtheorem{thm}{Theorem}
\newtheorem{remk}{Remark}
\newtheorem{cor}{Corollary}
\newtheorem{lem}{Lemma}

\newtheorem{defn}{Definition}

\usepackage[final]{graphicx}
\usepackage[reqno]{amsmath}
\usepackage{amssymb}
\usepackage{subfig}
\usepackage{epstopdf}
\usepackage{bm}
\usepackage{algorithm}
\usepackage{algorithmicx}
\usepackage{algpseudocode}
\usepackage[usenames, dvipsnames]{color}
\usepackage{bbm}

\algdef{SE}[DOWHILE]{Do}{doWhile}{\algorithmicdo}[1]{\algorithmicwhile\ #1}

\begin{document}

\sloppy
\title{Degrees of Freedom in Wireless Interference Networks with Cooperative Transmission and Backhaul Load Constraints\footnote{Parts of this work were presented at the International Symposium on Information Theory (ISIT) in 2014~\cite{isit14} and 2015~\cite{isit15}.}}

\author{
        Meghana~Bande,~\IEEEmembership{Student~Member,~IEEE,}
        Aly~El~Gamal,~\IEEEmembership{Member,~IEEE,}
        and~Venugopal~V.~Veeravalli,~\IEEEmembership{Fellow,~IEEE}
\thanks{A. {El Gamal} is with the Department of Electrical and Computer Engineering, Purdue University, IN 47907 USA (e-mail: elgamala@purdue.edu).}
\thanks{M. Bande and V. V. Veeravalli are with the Coordinated Science Laboratory and the Department of Electrical and Computer Engineering, University of Illinois
at Urbana-Champaign, Urbana, IL 61801 USA (e-mail: mbande2@illinois.edu, vvv@illinois.edu).}
\thanks{This research was supported in part by the US NSF WIFiUS Program under grant number 1457168 and the US
NSF SpecEES under grant number 1730882, through the University of Illinois at Urbana-Champaign, and in part by Huawei Innovation Research Program (HIRP) under grant number 301689, through Purdue University.}}
%s\thanks{Manuscript received April 19, 2005; revised August 26, 2015.}}

\maketitle

\begin{abstract}

Degrees of freedom (DoF) gains are studied in wireless networks with cooperative transmission under a \emph{backhaul load} constraint that limits the average number of messages that can be delivered from a centralized controller to basestation transmitters. The backhaul load is defined as the sum of all the messages available at all the transmitters per channel use, normalized by the number of users. For Wyner's linear interference network, where each transmitter is connected to the receiver having the same index as well as one succeeding receiver, the per user DoF is characterized and the optimal scheme is presented. %When the backhaul load is constrained to an integer $B$, the asymptotic per user DoF is shown to equal $\frac{4B-1}{4B}$.
 Furthermore, it is shown that the optimal assignment of messages to transmitters is asymmetric and satisfies a local cooperation constraint, and that the optimal coding scheme relies only on one-shot cooperative zero-forcing transmit beamforming. 
Using insights from the analysis of Wyner's linear interference network, the results are extended to the more practical hexagonal sectored cellular network, and coding schemes based on cooperative zero-forcing are shown to deliver significant DoF gains.
% It is also shown that for $B<5$, the asymptotic per user DoF for zero-forcing schemes is upper bounded by $\frac{5+B}{10}$.} 
%For the case of no additional backhaul load, a per user DoF of $\frac{1}{2}$ can be achieved. It is also shown that if each message can only be available at a single transmitter (no cooperative transmission), $\frac{1}{2}$ is an upper bound on the per user DoF, and $\frac{2}{5}$ is an upper bound on the per user DoF achievable through one-shot interference avoidance. 
It is established that by allowing for cooperative transmission and a flexible message assignment that is constrained only by an average backhaul load, one can deliver the rate gains promised by \emph{information-theoretic upper bounds} with practical one-shot schemes that incur little or no additional load on the backhaul. Finally, useful upper bounds on the per user DoF for schemes based on cooperative zero-forcing are presented for lower values of the average backhaul load constraint, and an optimization framework is formulated for the general converse problem.

\end{abstract}

\section{Introduction}

Managing interference in wireless networks has emerged as a challenging and important task over the past decade. 
We explore the potential degrees of freedom gains of cooperative transmission in wireless networks through different models for the interference, and under average backhaul load constraints. In particular, we show that cooperative transmission can be used to achieve significant DoF gains without requiring extra backhaul capacity.

We begin by studying the degrees of freedom (DoF) in Wyner's linear interference network, introduced in~\cite{Wyner}, where interference is modeled by assuming that the transmission of each transmitter is heard only by the receiver that has the same index as well as one succeeding transmitter. Wyner's model, while being simple, allows us to obtain rigorous conclusions about the optimal schemes for interference management. Further, as we show in this work, the insights obtained through analyzing linear networks such as Wyner's network can often be carried over to more complex network models that better approximate practical wireless networks.

Our focus on the DoF criterion is justified by the fact that it is useful to capture roughly the available capacity as a fraction of the capacity of an interference-free network consisting of point-to-point links. Two major advantages of the DoF criterion are as follows: 
(i) it is easy to analyze, and in many cases, the problem of finding an information theoretic upper bound or converse reduces to a straightforward combinatorial problem
; and (ii) it  captures the effect of interference, while circumventing the difficulties in analysis introduced by the additive Gaussian noise at the receivers. The DoF of a point-to-point link with white Gaussian noise is unity, and this is the reference benchmark for any given user's rate in an interference network, i.e., the per user DoF is at most one.

The DoF gain offered by cooperative transmission\footnote{Also called Coordinated Multi-Point (CoMP) transmission~\cite{CoMP-book}.} in Wyner's linear interference networks was studied in~\cite{Lapidoth-Shamai-Wigger-ISIT07}, for the special case where each message is available at the transmitter with the same index as well as $M-1$ succeeding transmitters. The asymptotic limit of the per user DoF as the number of users goes to infinity was shown to be $\frac{M}{M+1}$. An asymptotic per user DoF of $\frac{2M-1}{2M}$ was achieved using a smarter message assignment in~\cite{Shamai-Wigger-ISIT11}. In the proposed scheme of~\cite{Shamai-Wigger-ISIT11}, each message is assigned to the transmitter with the same index as well as $M-1$ other transmitters. However, unlike the assignment of~\cite{Lapidoth-Shamai-Wigger-ISIT07}, in~\cite{Shamai-Wigger-ISIT11} the choice of the $M-1$ other transmitters is not simply the succeeding $M-1$ transmitters. In~\cite{ElGamal-Annapureddy-Veeravalli-arXiv12}, it is shown that under a maximum transmit set size contraint constraint that limits the number of transmitters at which each message can be available by $M$, the asymptotic per user DoF is $\frac{2M}{2M+1}$ and is achieved by a \emph{flexible} assignment of messages to transmitters where it is not necessary to assign each message to the transmitter with the same index.
The DoF gains discussed in~\cite{Lapidoth-Shamai-Wigger-ISIT07},~\cite{Shamai-Wigger-ISIT11} and~\cite{ElGamal-Annapureddy-Veeravalli-arXiv12} are achieved by a simple signaling scheme that relies only on zero-forcing transmit beamforming.
 
The maximum transmit set size constraint of $M$ is not met tightly for all messages in the optimal message assignment scheme presented in~\cite{ElGamal-Annapureddy-Veeravalli-arXiv12}. In this work, we therefore consider a cooperation constraint that is more general and relevant to many scenarios of practical significance. In particular, we define the \emph{backhaul load} constraint $B$ as the ratio between the sum of the transmit set sizes for all the messages and the number of users. In other words, we allow the transmit set size to vary across the messages, while maintaining a constraint on the average transmit set size of $B$. We establish in this paper that the asymptotic per user DoF in this new setting is $\frac{4B-1}{4B}$, which is larger than the per user DoF of $\frac{2M}{2M+1}$ obtained with the more stringent per message transmit set size constraint of $M=B$. The identified optimal scheme relies only on zero-forcing beamforming at the transmitters, and an asymmetric or unbalanced assignment of messages, with some messages being assigned to more than $B$ transmitters and others being assigned to fewer than $B$ transmitters. 

We apply these insights to the more practical hexagonal sectored cellular model. 
In particular, we show that with cooperative transmission that is based on zero-forcing beamforming with asymmetric assignment of messages under an integer backhaul load constraint of $B$, it is possible to achieve a per user DoF of $\frac{2B}{3B+1}$ (Theorem \ref{thm:main}). We also show that under restriction to zero-forcing schemes, the asymptotic per user DoF is upper bounded by $\frac{5+B}{10}$ for $B<5$ (Theorem \ref{zf:B}), and formulate the general problem of finding the maximum per user DoF under restriction to zero-forcing schemes as an optimization problem. We emphasize that a per user DoF of $\frac{1}{2}$ is achievable with simple zero-forcing schemes and $B=1$, i.e., with no additional backhaul load.
On the other hand, we show that if cooperative transmission is not allowed ($M=1$), then a per user DoF of  $\frac{1}{2}$ is the optimal value (Theorem \ref{thm:upper bound}), and cannot be obtained by simple interference avoidance schemes (Theorem \ref{thm:tdma}). 
This shows that simple one-shot zero-forcing beamforming combined with non-uniform message assignments can be used to achieve significant gains in the per user DoF, while maintaining a low average backhaul load.

\subsection{Related Work}
A major advance in the theoretical analysis of interference management in large wireless networks took place with the introduction of asymptotic interference alignment in~\cite{Cadambe-IA} (IA). IA beamforming relies on signaling over a number of time slots ({\em symbol extension}) that goes to infinity in order to enable the achievability of a per user DoF of $\frac{1}{2}$ in a fully connected interference network. However, the gains offered by IA are considered to be infeasible in practice, and a major reason for the infeasability is the excessive requirement on the length of symbol extension, which would lead to impractical delays. An important aspect of this work is that we show that the promised gains of interference alignment can be achieved with one-shot coding schemes that do not require symbol extension, if we consider more practical network models than the fully connected model and allow for cooperative transmission, even without requiring additional overall load on the supporting backhaul.

Degrees of freedom gains in the hexagonal cellular downlink using cooperative transmission was considered in \cite{ntranos2014uplink}, where the transmitting basestations cooperate by exchanging quantized dirty paper coded signals. However, implementing such a scheme can face practical challenges as each transmitter gets its message only after a series of preceding transmitters have encoded their messages; this will require either significant delay or coding over multiple time slots. Further, under this setting, the only way for messages to be delivered to transmitters through a centralized controller, is for the controller to be aware of the channel state information. 

\subsection{Document Organization}
We describe the system model in Section~\ref{sec:systemmodel}. In Section~\ref{sec:proof}, we outline the arguments that we use throughout the paper for deriving degrees of freedom lower and upper bounds. We then characterize the degrees of freedom for the Wyner linear network in Section \ref{sec:Wyner}. We extend the results to the 
%linear locally connected channel model in Section \ref{sec:local}, and to the
% two-dimensional Wyner network are discussed in Section \ref{sec:2dWyner} and for the 
 hexagonal cellular network in Section \ref{sec:hexa}. Finally, we provide concluding remarks in Section~\ref{sec:conclusion}.

\section{System Model and Notation}\label{sec:systemmodel}
We use the standard model for the $K$-user interference channel with single-antenna transmitters and receivers,
%\begin{equation}
%Y_i(t) = \sum_{j=1}^{K} H_{i,j}(t) X_j(t) + Z_i(t),
%\end{equation}
%where $t$ is the time index, $X_j(t)$ is the transmitted signal of transmitter $j$, $Y_i(t)$ is the received signal at receiver $i$, $Z_i(t)$ is the zero mean unit variance Gaussian noise at receiver $i$, and $H_{i,j}(t)$ is the channel coefficient from transmitter $j$ to receiver $i$ over the time slot $t$. We remove the time index in the rest of the paper for brevity unless it is needed. For any set ${\cal A} \subseteq [K]$, we use the abbreviations $X_{\cal A}$, $Y_{\cal A}$, and $Z_{\cal A}$ to denote the sets $\left\{X_i, i\in {\cal A}\right\}$, $\left\{Y_i, i\in {\cal A}\right\}$, and $\left\{Z_i, i\in {\cal A}\right\}$, respectively. Finally, we use $[K]$ to denote the set $\{1,2,\ldots,K\}$, and use $\phi$ to denote the empty set.
%
%Consider a network with $K$ users. The signal $Y_{i}$ at
%receiver $i$ is given by

%\begin{equation}\label{signal}
%Y_{i}=H_{i,i}X_{i}+\sum_{j\in {\cal N}_{i}}H_{i,j}X_{j}+Z_{i}
%\end{equation}

\begin{equation}\label{signal}
Y_{i}=\sum_{j\in {\cal N}_{i}}H_{i,j}X_{j}+Z_{i}
\end{equation}

where $X_{j}$ denotes the signal transmitted by transmitter $j$
under an average transmit power constraint, $Z_{i}$ denotes the additive white
Gaussian noise at receiver $i$, $H_{i,j}$ denotes the channel
gain coefficient from transmitter $j$ to receiver $i$, and ${\cal N}_{i}$ denotes the
the set of transmitters that can be heard at receiver $i$ (neighbors in the connectivity graph including itself).
All channel coefficients that are not identically zero are assumed to be drawn from a continuous joint distribution. Finally, it is assumed that global channel state information is available at all transmitters and receivers. 

\subsection{Linear Interference Networks}
\subsubsection{Wyner assymmetric model}
In this channel model, each transmitter is connected to its corresponding receiver as well as one following receiver, and the last transmitter is only connected to its corresponding receiver. More precisely,

\begin{equation}\label{eq:channel}
H_{i,j} \equiv 0 \text { iff } i \notin \{j,j+1\},\forall i,j \in [K],
\end{equation}

where $[K]$ denotes the set $\{1,2, \ldots, K\}$.

\subsubsection{Locally connected channels}
 This is a more general linear network defined in~\cite{ElGamal-Annapureddy-Veeravalli-arXiv12}, where each receiver sees interference from $L$ neighboring transmitters. More precisely, for the following channel model,
%\begin{eqnarray}\label{eq:general_channel}
%&&H_{i,j} \text{ is not identically } 0, \nonumber\\&&\text { if and only if } i \in \left[j- \left \lfloor \frac{L}{2} \right \rfloor , j+ \left \lceil \frac{L}{2} \right \rceil \right].
%\end{eqnarray}

\begin{equation}\label{eq:general_channel}
H_{i,j} \equiv 0 \text { iff } i \notin \left[j- \left \lfloor \frac{L}{2} \right \rfloor , j+ \left \lceil \frac{L}{2} \right \rceil \right],\forall i,j \in [K].
\end{equation}

%\textcolor{blue}{
%\subsection{Two-dimensional Wyner network}\label{sub:two_d}
%Consider the two-dimensional network depicted in Figure \ref{fig_twod} where each transmitter is connected to four cell edge receivers. The precise channel model for a $K$-user channel is as follows,
%%\begin{eqnarray}\label{eq:twodim_channel}
%%&&H_{i,j} \text{ is not identically } 0, \text { if and only if }\nonumber\\&& i \in \left\{j,j+1,j+\left \lfloor \sqrt{K} \right \rfloor, j+\left \lfloor \sqrt{K} \right \rfloor+1\right\}.\nonumber
%%\end{eqnarray}
%\begin{eqnarray}\label{eq:twodim_channel}
%&&H_{i,j} \text{ is not identically } 0, \text { if and only if }\nonumber\\&& i \in \left\{j,j+1,j+\left \lfloor \sqrt{K} \right \rfloor, j+\left \lfloor \sqrt{K} \right \rfloor+1\right\}.\nonumber\\
%\end{eqnarray}}
%\begin{figure}[htb]
%  \centering
%\includegraphics[height=0.3\textwidth]{two-dim.eps}              
%\protect \caption{Channel model for the two dimensional interference network.}
%\label{fig_twod}
%\end{figure}

\subsection{Hexagonal Cellular Network}\label{sub:cell}
\begin{figure}[htb]
\centering
\includegraphics[scale=0.6]{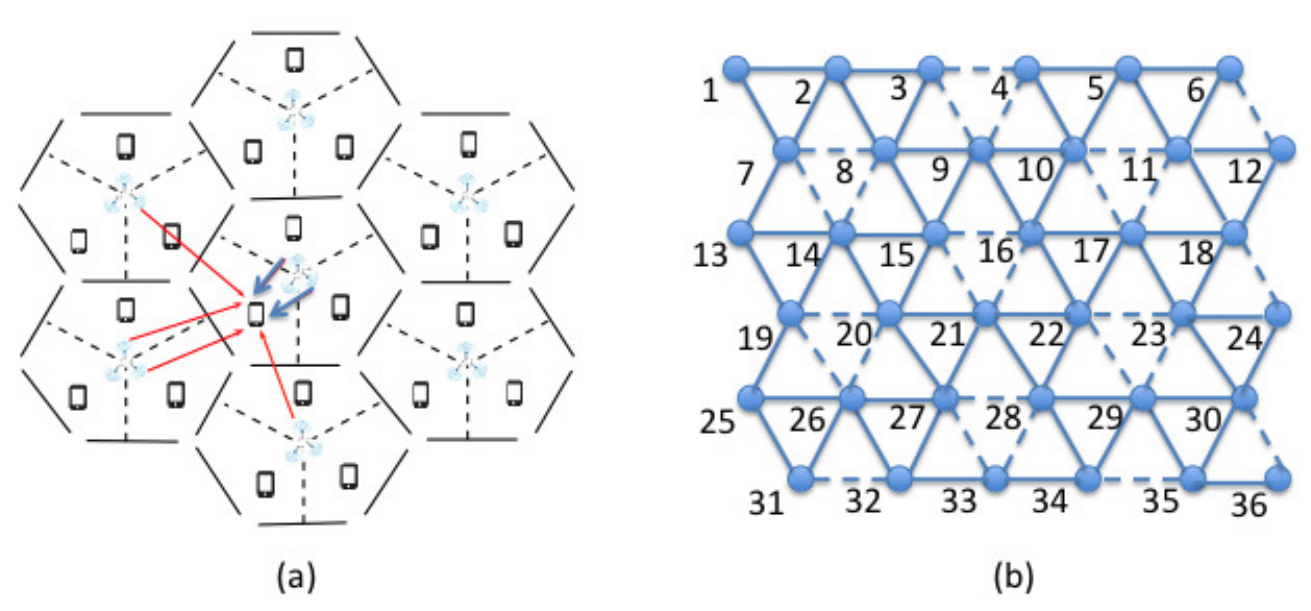}
\caption{(a) Cellular network and (b) connectivity graph. The dotted lines in (b) represent the interference between sectors belonging to the same cell.}
\label{Cellular Network}
\end{figure}

This is a sectored $K$ user cellular network with three sectors per
cell as shown in Figure \ref{Cellular Network}(a). We assume a local interference model, where the interference at each receiver is only due to the basestations in the neighboring sectors in adjacent cells. It is assumed that the sectors belonging to the same cell do not interfere with each other, the justification being that the interference power due to sectors in the same cell is usually far lower than the interference from out-of-cell users located in the sector's line of sight. 

\subsubsection{Connectivity graph}\label{subsec:con_graph}
The cellular model is represented by an undirected connectivity graph
$G(V,E)$ shown in Figure \ref{Cellular Network}(b) where each vertex $u\in V$ corresponds
to a transmitter-receiver pair. For any node $a$, the transmitter, receiver and intended message (word) corresponding to the node are denoted by $T_{a}$, $R_{a}$ and $W_{a}$, respectively. An edge $e\in E$ between two vertices  $u,v\in V$ corresponds
to a channel existing between the transmitter at $u$ and the receiver at $v$, and vice-versa. The dotted lines denote interference between sectors that belong to the same cell, and is ignored in our model. For any node $a$, ${\cal N}_a$ denotes the set of nodes adjacent to node $a$ and that includes node $a$. To simplify the presentation, without much loss of generality, we consider only $K-$user networks where $\sqrt{K}$ is an integer, and nodes are numbered as in Figure~\ref{Cellular Network}(b). (In the figure, $\sqrt{K}=6$). Since we are studying the performance in the asymptotic limit of the number of users, the assumption is not restrictive.

\begin{figure}[htb]
\centering
\includegraphics[scale=0.6]{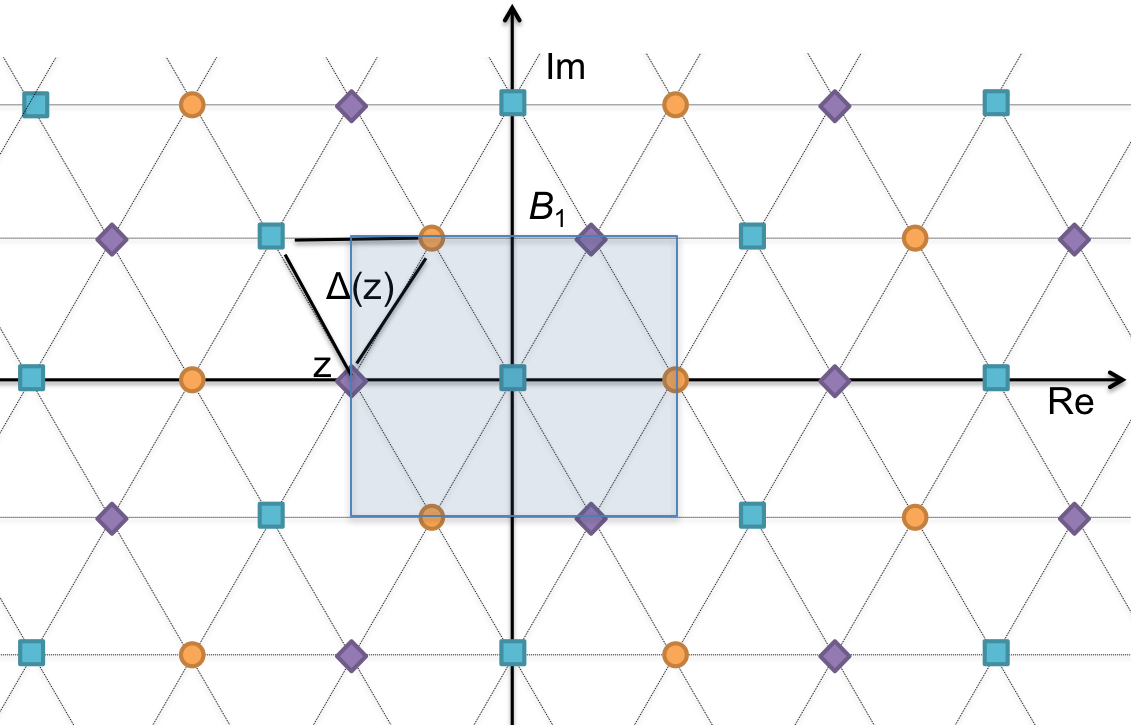}
\caption{The cellular network is represented by Eisenstein integers, partitioned into three cosets $\Omega_{sq},\Omega_{cir},\Omega_{dia}$ represented by square, circle and diamond shaped nodes respectively. For any node $z$, $\Delta(z)$ represents the edges between the nodes, $z$, $z+\omega$ and $z+\omega+1$. $\mathbb{B}_r$ is illustrated in the figure for $r=1$. The seven nodes lying on or within the rectangle belong to the set $\mathbb{B}_1$.}
\label{eisen}
\end{figure}

We formally define the connectivity graph $G(V,E)$ using Eisenstein integers similar to \cite{ntranos2014uplink}.
\begin{defn} (Eisenstein integers) :
Eisenstein integers $\mathbb{Z}[\omega]$ are numbers of the form $a+b\omega$ where $\omega = \frac{1}{2}(-1+\imath\sqrt{3}) $ and $a,b \in \mathbb{Z}$, where $\imath=\sqrt{-1}$.
\end{defn}
Let $\mathbb{B}_r = \{z\in \mathbb{C}: |\text{Re}(z)|\leq r, |\text{Im}(z)|\leq \frac{\sqrt{3}r}{2}\}$. The set $\mathbb{B}_r$ denotes the Eisenstein integers enclosed in the rectangle centered at the origin with the real part bounded by $r$ and the imaginary part bounded by $\frac{\sqrt{3}r}{2}$. Consider the following one-to-one mapping $g:V\to \mathbb{Z}[\omega]\cap \mathbb{B}_r  $ between vertices of the graph and Eisenstein integers.
For each $v\in V$, $g(v)$ denotes the corresponding vertex in the Eisenstein graph. Note that 
\begin{equation}\label{eq:vert}
V = \{g^{-1}(z):z \in \mathbb{Z}[\omega]\cap \mathbb{B}_r \}.
\end{equation}
Consider the function $f(a+b\omega) =  (a+b) \text{ mod 3}$. This partitions the space of Eisenstein integers into three cosets represented by $\Omega_{sq},\Omega_{cir},\Omega_{dia}$ corresponding to $f(z)=0$, $f(z)=1$ and $f(z)=2$ for all $z\in \mathbb{Z}[\omega]$. The subscripts of $\Omega_{sq},\Omega_{cir},\Omega_{dia}$ correspond to the squares, circles and diamonds which are used to represent the respective cosets in Figure \ref{eisen}. 
%For any $z \in \mathbb{Z}[\omega]\cap \mathbb{B}_r$, we define the following set of edges  $\Delta(z)$ forming a triangle between the vertices $z,z+\omega$ and $z+\omega+1$,
%\[
%\Delta(z) = \{(z,z+\omega),(z,z+\omega+1),(z+\omega,z+\omega+1)\}.
%\]

For any $z \in \mathbb{Z}[\omega]\cap \mathbb{B}_r$, we define the following triangle  $\Delta(z)$ ,
\[
\Delta(z) = \{z,z+\omega,z+\omega+1\},
\]
and the edges incident to the vertices of $\Delta(z)$ are denoted by ${\cal E}(\Delta(z))$ as follows,
\[
{\cal E}(\Delta(z)) = \{(z,z+\omega),(z,z+\omega+1),(z+\omega,z+\omega+1)\}.
\]

If we consider the edges ${\cal E}(\Delta(z))$, where $z\in \Omega_{sq}$, each node is incident to exactly two edges and by removing these edges, we have the connectivity graph in Figure \ref{proof_pic}, a proper representation of the hexagonal cellular network with no intra-cell interference. 
More precisely, without loss of generality, let ${\cal E}(\Delta(z))$, where $z\in \Omega_{sq}$, correspond to the intra-cell interference. Then since we ignore intra-cell interference in our model, we define the set of interfering edges in the graph as
\begin{equation}\label{eq:edges}
E = \{ (u,v): u,v \in V \text{and } (g(u),g(v)) \in {\cal E}(D)\},
\end{equation}

where,
\[
D = \{ \Delta(z): z \in \{\Omega_{cir}\cup \Omega_{dia}\}\}.
\]
Thus the interference graph is $G(V,E)$ where $V$ is given by \eqref{eq:vert} and the set of edges $E$ is given by \eqref{eq:edges}.

\subsection{Message Assignment}
For each $i \in [K]$, let $W_i$ be the message intended for receiver $i$, and ${\cal T}_i \subseteq [K]$ be the transmit set of receiver $i$, i.e., those transmitters with the knowledge of $W_i$. The transmitters in ${\cal T}_i$ cooperatively transmit the message $W_i$ to the receiver $i$. A particular message assignment is
denoted by $\{{\cal T}_{i}\}_{i\in [K]}$. For a particular message assignment, $M$ denotes the maximum transmit set size and $B$ denotes the backhaul load or the average transmit set size,
\begin{equation}\label{M}
M=\underset{i}{\text{max}}|{\cal T}_{i}|,
\end{equation}
\begin{equation}\label{B}
B=\frac{\sum_{i=1}^K|{\cal T}_{i}|}{K}.
\end{equation}

In this work, we allow for flexible association of messages, i.e., 
we only restrict the size of transmit sets, without constraints on the specific set of transmitters that each message is assigned to. The case $M=1$ corresponds to the case of no cooperation, but with possibly a flexible association of cells. The case $B=1$ corresponds to an average backhaul load of one message per transmitter, i.e., no extra backhaul load due to cooperation.

\subsection{Local Cooperation}\label{sec:localcooperation}

We say that the local cooperation constraint is satisfied, if and only if there exists a function $r(K)$ such that $r(K)=o(K)$, and for every $K\in{\bm Z}^{+}$, the transmit sets used for the $K$-user channel satisfy the following:
\begin{equation}\label{eq:localcooperation}
{\cal T}_{i} \subseteq \{i-r(K),i-r(K)+1,\ldots,i+r(K)\}, \forall i\in[K].
\end{equation}

%\subsection{Zero-forcing (Interference Avoidance) Schemes}\label{sec:zf}
%We consider in this work the class of \textit{interference avoidance} schemes, where each message is either not transmitted or allocated one degree of freedom.
%Accordingly, every receiver is either active or inactive. An active receiver does not observe any interfering signals.

\subsection{Zero-forcing schemes}\label{sec:zf}

We consider in this work the class of \textit{zero-forcing} schemes, where each message is either not transmitted or allocated one degree of freedom.
Accordingly, every receiver is either active or inactive. An active receiver does not observe any interfering signals. For the case of no-cooperation i.e., $M=1$, we refer to these schemes as interference avoidance schemes. The case where $M \geq 2$ corresponds to the scenario where cooperative zero-forcing can be used.

For any zero-forcing scheme, the transmit signal at the $j^{\mathrm{th}}$ transmitter is given by,
\begin{equation}\label{eq:linear-precoding}
X_j = \sum_{i: j \in {\cal T}_i} X_{j,i},
\end{equation} 
where $X_{j,i}$ depends only on message $W_i$. Further, each message is either not transmitted or allocated one degree of freedom. More precisely, let $\tilde{Y}_j=Y_j-Z_j, \forall j\in[K]$. Then, in addition to the constraint in~\eqref{eq:linear-precoding}, it is either case that the mutual information $I(\tilde{Y}_j;W_j)=0$ or it is the case that $W_j$ completely determines $\tilde{Y}_j$. Note that $\tilde{Y}_j$ can be determined from $W_j$ for the case where user $j$ enjoys interference-free communication, and $I(W_j;\tilde{Y}_j)=0$ for the other case where $W_j$ is not transmitted. We say that the $j^{\mathrm{th}}$ receiver is \emph{active} if and only if $I(\tilde{Y}_j;W_j)>0$. Note that using zero-forcing transmit beamforming, if the $j^{\mathrm{th}}$ receiver is active, then $I(W_i;Y_j)=0, \forall i \neq j$. Finally, we say that the $j^{\mathrm{th}}$ transmitter is \emph{active} if $I\left(X_j;\{W_i: j\in {\cal T}_i\}\right)>0$.

Without loss in generality, we assume that it has to be the case that if a message $W_i$ is available at transmitter $j$, i.e., $j \in {\cal T}_i$, then the message contributes to the corresponding transmit signal, i.e., $I(W_i, X_{j,i})>0$. Otherwise, the message assignment could be removed without affecting the sum rate.

\subsection{Degrees of Freedom}\label{subset:dof}
Let $P$ be the average transmit power constraint at each transmitter, and let ${\cal W}_i$ denote the alphabet for message $W_i$. Then the rates $R_i(P) = \frac{\log|{\cal W}_i|}{n}$ are achievable if the decoding error probabilities of all messages can be simultaneously made arbitrarily small for a large enough coding block length $n$, and this holds for almost all channel realizations. The degrees of freedom (DoF) $d_i, i\in[K],$ is defined as $d_i=\lim_{P \rightarrow \infty} \frac{R_i(P)}{\log P}$. The DoF region ${\cal D}$ is the closure of the set of all achievable DoF tuples. The total DoF ($\eta$) is the maximum value of the sum of the achievable degrees of freedom, $\eta=\max_{\cal D} \sum_{i \in [K]} d_i$.

 For a $K$-user channel, we define $\eta(K,M)$ and $\eta^{\text{avg}}(K,B)$ as the maximum achievable $\eta$ over all possible message assignments satisfying the constraints (\ref{M}) and (\ref{B}) respectively. We define the following asymptotic
quantities which capture how $\eta$ scales with $K$. 
\begin{equation}
\tau(M)=\underset{K\rightarrow\infty}{\text{lim}}\frac{\eta(K,M)}{K},
\end{equation}
\begin{equation}
\tau^{\text{avg}}(B)=\underset{K\rightarrow\infty}{\text{lim}}\frac{\eta^{\text{avg}}(K,B)}{K}.
\end{equation}

We use the superscript $\text{zf}$ to indicate a further restriction to zero-forcing schemes. Finally, we denote the DoF and asymptotic per user DoF 
%for locally connected networks with connectivity parameter $L$ with subscript $L$, and 
for the hexagonal cellular network with subscript $c$.

\section{Proof Techniques}\label{sec:proof}
Before discussing the results we have for the above introduced network models, we provide in this section a brief overview of the main arguments used in the achievability and converse proofs throughout this work. 

\subsection{Message Assignments and Coding Schemes}

For the considered system model, a proof of achievability involves a choice of assigning messages to transmitters, and a transmission scheme that indicates coding and scheduling decisions. We employ \emph{interference-aware} message assignments that divide the network into subnetworks of optimal size, where each subnetwork consists of a fixed number of transmitter-receiver pairs. The messages destined for receivers in a subnetwork can only be assigned to transmitters within the same subnetwork. Hence, all of our message assignments satisfy the local cooperation constraint in~\eqref{eq:localcooperation}. We then employ a zero-forcing scheme that guarantees complete interference cancellation for all \emph{active} receivers within a subnetwork. Because we assume that all channel coefficients are drawn from a continuous joint distribution, thereby ensuring that the probability of any specific realization is zero, cancelling a message's interference at a number $n$ of undesired receivers would require assigning this message at $n$ transmitters other than the transmitter delivering the message. Hence, the backhaul load constraint induces a constraint on the number of receivers at which each message's interference can be canceled, and accordingly, a constraint on the size of each subnetwork. 

In~\cite{ElGamal-Annapureddy-Veeravalli-arXiv12}, a cooperation constraint the limits the maximum transmit set size was considered. The asymptotic per user DoF was then characterized for Wyner's linear asymmetric network and achievable per user DoF values were presented for other network models. Here, we observe that we can employ the results obtained in~\cite{ElGamal-Annapureddy-Veeravalli-arXiv12} by using convex combinations of the schemes that are optimal under a maximum transmit set size constraint, in order to obtain a scheme that is optimal under the considered average transmit set size constraint. By a \emph{convex combintation} of schemes, we refer to employing each of the schemes in a part of the network that consists of a number of successive transmitter-receiver pairs, and that number equals a fraction of the total number of users, with the sum of these fractions equaling unity. 

It is interesting to observe that local cooperation combined with one-shot zero-forcing schemes can be used to achieve significant scalable DoF gains in large networks. Further, these gains can be achieved with no or minimal extra load on the backhaul, since deactivating few transmitters not only helps with avoiding interference and splitting the network into small subnetworks, but also releases backhaul resources that could be used to assign active messages at more than one transmitter. Finally, it is worth noting that even though we only capture the sum rate through the asymptotic per user DoF criterion, fairness between the users could be achieved through a \emph{fractional reuse} mechanism where user indices are shifted across time or frequency slots.

\subsection{Converse Proofs}
The converse proofs presented in this work rely on two fundamental results, that were proved in~\cite{ElGamal-Annapureddy-Veeravalli-arXiv12}. First, we use Lemma $4$ from~\cite{ElGamal-Annapureddy-Veeravalli-arXiv12}, which we restate below. For any set of receiver indices ${\cal A} \subseteq [K]$, define $U_{\cal A}$ as the set of indices of transmitters that exclusively carry the messages for the receivers in ${\cal A}$, and its complement $\bar{U}_{\cal A}$. More precisely, $\bar{U}_{\cal A} = \cup_{i \notin {\cal A}} {\cal T}_i$.

\begin{lem} [\protect{\cite[Lemma 4]{ElGamal-Annapureddy-Veeravalli-arXiv12}}] \label{lem:dofouterbound}
If there exists a set ${\cal A}\subseteq [K]$, a function $f_1$, and a function $f_2$ whose definition does not depend on the transmit power constraint $P$, and $f_1\left(Y_{\cal A},X_{U_{\cal A}}\right)=X_{\bar{U}_{\cal A}}+f_2(Z_{\cal A})$, then the sum DoF $\eta \leq |{\cal A}|$.
\end{lem}

What Lemma~\ref{lem:dofouterbound} implies is that if there is a centralized decoder that has access to all the received signals $Y_{\cal A}$, and a reliable communication scheme is used, then this decoder would be able to decode all the $K$ messages, and hence, the DoF is bounded by the number of signals used for decoding $|{\cal A}|$. First, the centralized decoder would be able to decode the messages $W_{\cal A}$ because we assumed that the communication scheme is reliable. The transmit signals $X_{U_{\cal A}}$ can then be reconstructed, since their reconstruction solely relies on the messages $W_{\cal A}$. Using the functions $f_1$ and $f_2$ in the statement of the lemma, the remaining transmit signals can then be reconstructed. Finally, using the knowledge of all the transmit signals, all the messages can be recovered with a vanishing error probability.

The second concept that we borrow from~\cite{ElGamal-Annapureddy-Veeravalli-arXiv12} is that of \emph{irreducible} message assignments. By reducibility, we refer to the possibility of removing a message assignment to one or more transmitters without affecting the sum rate, regardless of the choice of the coding scheme. A message assignment to a transmitter can only be useful either for delivering the message to its intended receiver, or for aiding in cancelling the message's interference at an unintended receiver. If it is guaranteed that neither functions can be achieved through a given assignment, then this assignment can be removed without affecting the sum rate. As a simple example for this argument, when cooperation is not allowed, any irreducible message assignment could have each message assigned only to one of the transmitters connected to its intended destinations.

When deriving a converse under the backhaul load (average transmit set size) constraint, we combine the above two concepts with combinatorial concentration inequalities that are based on the pigeonhole principle and allow us to infer from the backhaul load constraint facts about the existence of a number of messages whose transmit set sizes can be bounded by a maximum value, and have guarantees on that number. This step simplifies the combinatorial aspect of the problem by allowing us to restrict our attention to a narrower class of possible message assignments.

\section{Wyner Interference Network}\label{sec:Wyner}
In~\cite{ElGamal-Annapureddy-Veeravalli-arXiv12}, each transmit set size was bounded by a maximum transmit set size contraint constraint $M$, i.e., $|{\cal T}_i| \leq M, \forall i\in[K]$. The DoF achieving coding scheme was then characterized for every value of $M$. We now consider the problem with an average transmit set size constraint $B$ and show that the per user DoF $\tau^{\text{avg}}(B)$ can be achieved using a combination of the schemes that are characterized as optimal in~\cite{ElGamal-Annapureddy-Veeravalli-arXiv12} for the cases of $M=2B-1$ and $M=2B$. We now understand from this result that even though the maximum transmit set size constraint may not reflect a physical constraint, the solutions in~\cite{ElGamal-Annapureddy-Veeravalli-arXiv12} provide a set of tools that can be used to achieve the optimal per user DoF value under the more natural constraint on the total backhaul load that is considered in this work.

\subsection{Example: $B=1$}\label{sec:example}
\begin{figure}[htb]
\centering
\includegraphics[width=0.36\columnwidth]{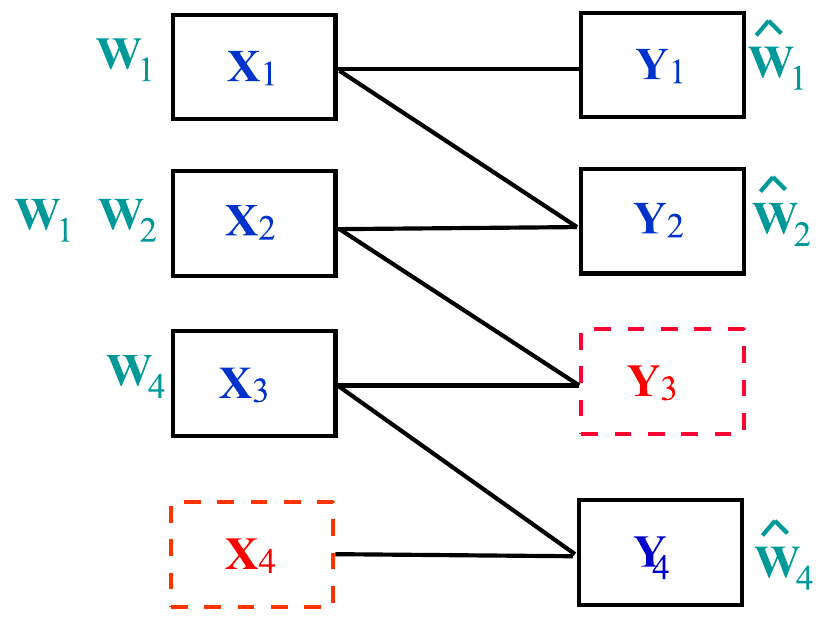}
\caption{Achieving $ \frac{3}{4}$ per user DoF with a backhaul load constraint $B=1$. The figure shows only signals corresponding to the first subnetwork in a general $K$-user network. The signals in the dashed boxes are deactivated.}
\label{fig:bone}
\end{figure}

Before introducing the main result, we illustrate through a simple example how the potential flexibility in the backhaul design according to the constraint in~(\ref{B}) can offer DoF gains over a traditional design where all messages are assigned to the same number of transmitters. We know from~\cite{ElGamal-Annapureddy-Veeravalli-arXiv12} that an asymptotic per user DoF greater than $\frac{2}{3}$ cannot be achieved through assigning each message to one transmitter. We now show that $\tau^{\text{avg}}(B=1)\geq\frac{3}{4}$, by allowing few messages to be available at more than one transmitter at the cost of not transmitting other messages. Consider the following assignment of the first four messages, ${\cal T}_1=\{1,2\}$, ${\cal T}_2=\{2\}$, ${\cal T}_3=\phi$, and ${\cal T}_4=\{3\}$. Note that the backhaul load constraint $B=1$ is respected, because $\frac{\sum_{i=1}^4 |{\cal T}_i|}{4}=1$. Message $W_1$ is transmitted through $X_1$ to $Y_1$ without interference. Since  the channel state information is known at the second transmitter, the transmit beam for $W_1$ at $X_2$ can be designed to cancel the interference caused by $W_1$ at $Y_2$, and then $W_2$ can be transmitted through $X_2$ to $Y_2$ without interference. Finally, $W_4$ is transmitted through $X_3$ to $Y_4$ without interference. It follows that the sum DoF for the first four messages $\sum_{i=1}^4 d_i \geq 3$. Since the fourth transmitter is inactive, the subnetwork consisting of the first four users does not interfere with the rest of the network, and hence, we can see that $\tau^{\text{avg}}(B=1) \geq \frac{3}{4}$ through a similar assignment of messages in each consecutive $4$-user subnetwork. We illustrate this example in Figure~\ref{fig:bone}.

\subsection{Main Result}\label{sec:main}
We now characterize the asymptotic per user DoF $\tau^{\text{avg}}(B)$ for any integer value of the backhaul load constraint.
\begin{thm}\label{thm:main_result}
The asymptotic per user DoF $\tau^{\text{avg}}(B)$ is given by,
\begin{equation}\label{eq:main_result}
\tau^{\text{avg}}(B)=\frac{4B-1}{4B}, \forall B\in{\bf Z}^+.
\end{equation}
\end{thm}
\begin{IEEEproof}
We provide the proof for the lower bound here; the proof of the upper bound is presented in the appendix. 
%\subsubsection{Coding Scheme}\label{sec:ib}
We treat the network as a set of subnetworks, each consisting of consecutive $4B$ transceivers. The last transmitter of each subnetwork is deactivated to eliminate {\em inter-subnetwork} interference. It then suffices to show that a DoF of $4B-1$ can be achieved in each subnetwork. Without loss of generality, consider the cluster of users with indices in the set $[4B]$. This is illustrated for $B=2$ in Figure~\ref{fig:btwo}. We define the following subsets of $[4B]$,
\begin{eqnarray*}
{\cal S}_1 &=& [2B],
\\{\cal S}_2 &=& \{2B+2,2B+3,\ldots,4B\}.
\end{eqnarray*}
We next show that each user in ${\cal S}_1 \cup {\cal S}_2$ achieves one degree of freedom, while message $W_{2B+1}$ is not transmitted. Let the message assignments be as follows,\\

${\cal T}_{i}=
\begin{cases}
\{i,i+1,\ldots,2B\}, \quad &\forall i \in {\cal S}_1,\\
\{2B+1,2B+2,\ldots,i-1\},\quad &\forall  i \in {\cal S}_2,
\end{cases}$\\
and note that $\frac{\sum_{i=1}^{4B} |{\cal T}_i|}{4B}=B$, and hence, the constraint in~(\ref{B}) is satisfied. Now, due to the availability of channel state information at the transmitters, the transmit beams for message $W_i$ can be designed to cancel its effect at receivers with indices in the set ${\cal C}_i$, where,\\

${\cal C}_{i}=
\begin{cases}
\{i+1, i+2, \ldots,2B\},\quad  &\forall i \in {\cal S}_1,\\
\{2B+2,2B+3,\ldots,i-1\},\quad &\forall  i \in {\cal S}_2.
\end{cases}$\\

Note that both ${\cal C}_{2B}$ and ${\cal C}_{2B+2}$ equal the empty set, as both $W_{2B}$ and $W_{2B+2}$ do not contribute to interfering signals at receivers in the set $Y_{{\cal S}_1} \cup Y_{{\cal S}_2}$. The above scheme for $B=2$ is illustrated in Figure~\ref{fig:btwo}.
We conclude that each receiver whose index is in the set ${\cal S}_1\cup{\cal S}_2$ suffers only from Gaussian noise, thereby enjoying one degree of freedom. Since $|{\cal S}_1\cup{\cal S}_2|=4B-1$, it follows that $\sum_{i=1}^{4B} d_i \geq 4B-1$. Using a similar argument for each following subnetwork and noting that the last transmitter in each subnetwork is inactive to eliminate inter-subnetwork interference, we establish that $\tau^{\text{avg}}(B) \geq \frac{4B-1}{4B}$, thereby proving the lower bound in Theorem~\ref{thm:main_result}.

\begin{figure}[htb]
\centering
\includegraphics[width=0.36\columnwidth]{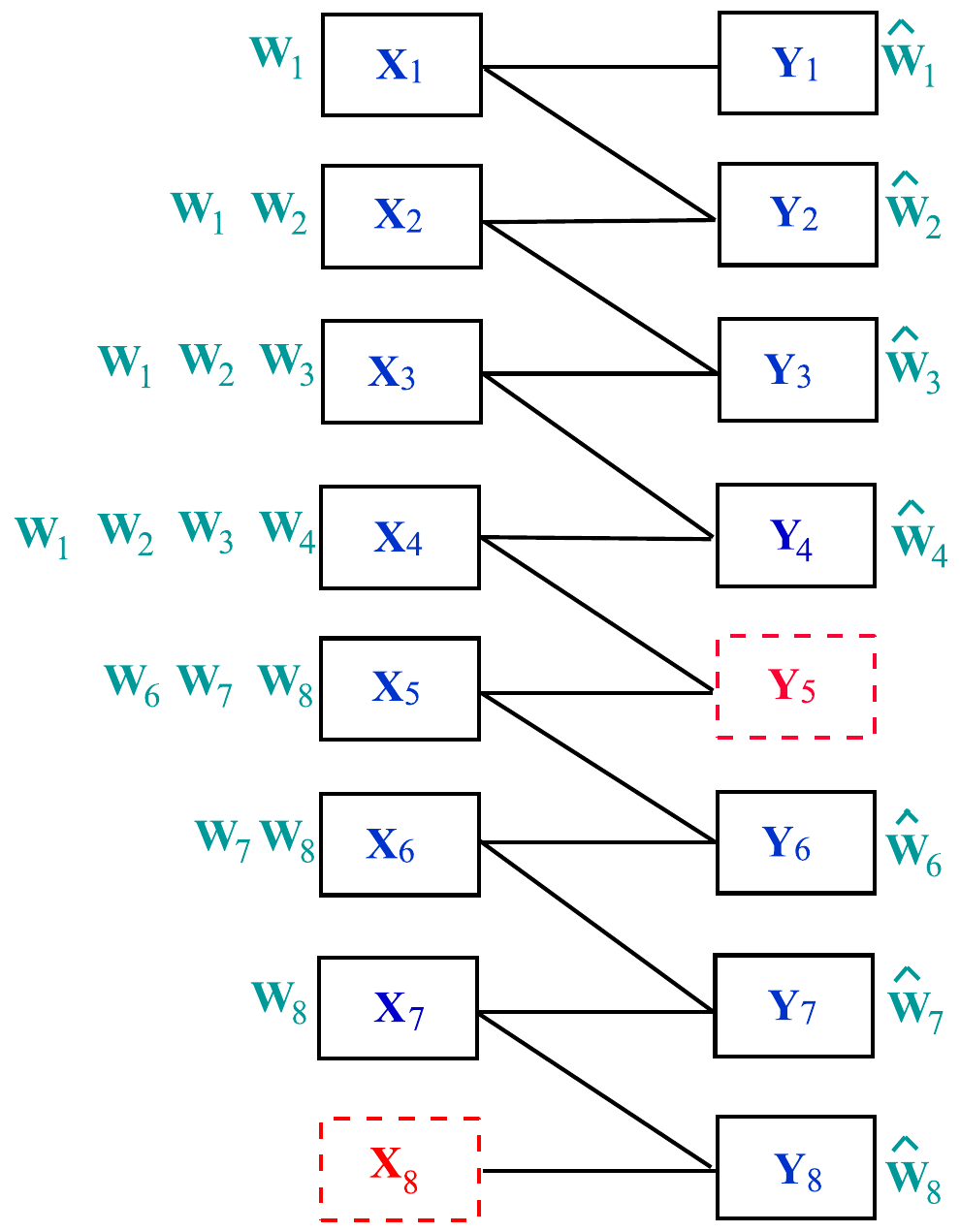}
\caption{Achieving $\frac{7}{8}$ per user DoF with a backhaul load constraint $B=2$. The figure shows only signals corresponding to the first subnetwork in a general $K$-user network. The signals in the dashed boxes are deactivated.}
\label{fig:btwo}
\end{figure} 

\end{IEEEproof}

We note that the local cooperation constraint of~\eqref{eq:localcooperation} is satisfied, when we use the illustrated message assignment. In other words, the network can be split into subnetworks, each of size $4B$, and the messages corresponding to users in a subnetwork can only be assigned to transmitters with indices in the same subnetwork. Few remarks are now in order.
%Local Cooperation

\begin{remk}
Although we assume availability of all channel coefficients at every transmitter in the network, the achievable schemes used  require only local channel state information.
\end{remk}
\begin{remk}
In the proposed achievable schemes, some messages are being sent interference free at the expense of other messages not being transmitted. Fairness can be maintained in the allocation of the available DoF over all users through fractional reuse by deactivating different sets of receivers in different sessions, e.g., in different time or frequency slots.
\end{remk}

\begin{remk}
The result of Theorem \ref{thm:main_result} can be achieved by a convex combination of the schemes in \cite{ElGamal-Annapureddy-Veeravalli-arXiv12} for $M=2B-1$ and $M=2B$ in the ratio $4B-1:4B+1$. Hence, even though the maximum transmit set size constraint that is used in the analysis of \cite{ElGamal-Annapureddy-Veeravalli-arXiv12} may not reflect a practical setting, but the obtained optimal schemes can be used to obtain the optimal scheme in the considered setting, where the more practical backhaul load constraint is considered.
\end{remk}
%\subsection{Discussion and Generalizations}\label{sec:discussion}

\begin{remk}\label{remk:generalL}
In a locally connected network with $L\geq 2$, the per user DoF for the case of no cooperation ($M=1$) is upper bounded by $\frac{1}{2}$. Further, with restriction to interference avoidance schemes, the DoF is upper bounded by $\frac{2}{2+L}$ ($\leq \frac{1}{2}$) \cite[Theorem 4]{ElGamal-Annapureddy-Veeravalli-arXiv12}.
 Similar to the proof of Theorem \ref{thm:main_result}, a per user DoF greater than or equal to $\frac{1}{2}$ can be achieved for $L= 2,\ldots,6$ by
a convex combination of the schemes described in \cite{ElGamal-Annapureddy-Veeravalli-arXiv12} under the maximum transmit set constraint $M$ using simple zero-forcing schemes, without the need for additional backhaul load, i.e., $B=1$. 
\end{remk}

\begin{remk}
The coding scheme used to prove the lower bound part of Theorem~\ref{thm:main_result} is similar to the scheme introduced in~\cite{Lapidoth-Shamai-Wigger-ISIT07}, as it relies on deactivating appropriately selected transmitters to maximize the number of interference free links. Mitigating interference among the remaining users is carried out through dedicating each assignment of a message to a transmitter either for message delivery at its intended destination, or for cancelling the message's interference at a single unintended destination. However, the scheme in~\cite{Lapidoth-Shamai-Wigger-ISIT07} relies on the dirty paper coding scheme introduced in~\cite{dirty-paper}, while here we are using zero-forcing transmit beamforming. We note that replacing zero-forcing transmit beamforming with dirty paper coding in the presented scheme would lead to the same DoF result.
\end{remk}

\section{Cellular network}\label{sec:hexa}

We have characterized the per user DoF for Wyner's linear interference network under the average  backhaul load constraint $B$. In this section, we investigate the per user DoF for the hexagonal sectored cellular model introduced in Section \ref{sub:cell}, using insights obtained from the analysis of Wyner's model. Our goal is to highlight the advantage of cooperative transmission that is based on flexible cell associations for cellular networks, by first showing that the asymptotic per user DoF is at most $\frac{1}{2}$ for the case when each message can be available at a single transmitter. Further, we show for this case that interference avoidance schemes can only be used to achieve an asymptotic per user DoF of at most $\frac{2}{5}$. On the other hand, when cooperative transmission is allowed, but the overall load on the backhaul is not increased ($B=1$), interference avoidance schemes can be used to achieve the $\frac{1}{2}$ asymptotic per user DoF value.

%\subsection{Main Results}\label{sec:main}

We first impose the maximum transmit set size constraint of $M=1$ in the network, i.e., a message of a cell edge mobile receiver can be assigned to any single basestation transmitter, thus leading to a flexible cell association in the cellular downlink. 

\begin{thm}\label{thm:upper bound}
For the considered hexagonal cellular network model, the following bound holds for the case of no-cooperation,
\[
\tau_c(M=1)\leq \frac{1}{2}.
\]
\end{thm}

\begin{IEEEproof} The proof is available in Section \ref{subsec:upper bound}.
\end{IEEEproof}

This shows that using a traditional approach for interference management, the maximum asymptotic per user DoF for the considered hexagonal cellular network model is $\frac{1}{2}$. Further, the only known way this DoF value can be approached is in the limit as the length of symbol extension goes to infinity as in the asymptotic interference alignment scheme of~\cite{Cadambe-IA}.

\subsection{Zero-forcing schemes}

In this section, we focus on cooperative zero-forcing, and interference avoidance which is a special case of zero-forcing schemes for $M=1$. We now introduce some additional notation. 
For each node $i \in [K]$, let $r_i$ indicate whether receiver $i$ is active or not, i.e., 
$r_{i}=\mathbbm{1}\{$Receiver $i$ is active$\}$,
where $\mathbbm{1}\{x\}$ is defined as,
$$
\mathbbm{1}\{x\}=
\begin{cases}
1, \text{ if $x$ is true} ,\\
0, \text{ otherwise}.
\end{cases}
$$
Similarly, for each node $i \in [K]$, let $t_i$ indicate whether transmitter $i$ is active or not, i.e., 
$t_{i}=\mathbbm{1}\{$Transmitter $i$ is active$\}$. We note that the sum DoF in the network is upper bounded by $\sum_{i\in[K]} r_i$. 
Consider the adjacency matrix $A$ of the connectivity graph $G$ of the network. The Edmond's matrix $D$ is defined as 
$$
D_{ij}=
\begin{cases}
x_{ij}, \text{ if } A_{ij}=1,\\
0, \text{ otherwise},
\end{cases}
$$

\begin{table}[htb]
\begin{center}
\begin{tabular}{ |c |c |} 
\hline
$t_i$& $\mathbbm{1}\{$Transmitter $i$ is active$\}$ \\ 
\hline
$r_i$ & $\mathbbm{1}\{$Receiver $i$ is active$\} $\\ 
\hline
${\cal N}_i$ & Set of nodes adjacent to node $i$ including node $i$\\ 
\hline
${\cal T}_{j}$ & Set of transmitters containing message $W_j$\\ 
\hline
$\rho_{j}$ & Fraction of users with messages available at exactly $j$ transmitters \\ 
\hline
${\cal V}_{\cal S}$ & Set of active receivers connected to transmitters in ${\cal S}$ \\ 
%\hline
%$d_{ij}$ & $\mathbbm{1}\{\text{Transmitter } i \text{ delivers message to receiver } j \text{ interference-free}\}$ \\ 
\hline
$D_{{\cal A},{\cal B}}$ & Edmond's matrix for bipartite graph with ${\cal A}$ and ${\cal B}$\\ 
\hline

\end{tabular}
\end{center}
\caption{Summary of notation used for zero-forcing bounds.}
\label{tab:notation}
\end{table}

where $x_{ij}$ are indeterminates. 
We note that a bipartite graph $G$ has a perfect matching if and only if the polynomial defined by the determinant $\mathrm{det}(D)$ is not identically zero i.e., $D$ has full rank. 
Let $D_{{\cal A},{\cal B}}$ denote the Edmond's matrix for the bipartite graph with ${\cal A}$ and ${\cal B}$ as the two partite sets, and any pair of vertices that have the same index are connected in the bipartite graph. 

For any set ${\cal S} \subseteq [K]$, we define ${\cal V}_{\cal S}$ as the set of indices of the active receivers connected to the transmitters with indices in ${\cal S}$.
 Then ${\cal V}_{\cal S}= \{k: k \in {\cal N}_i, i\in {\cal S} \text{ and } I(Y_k;W_k)>0\}$.

We need the following lemma, which is an extension of a lemma from \cite{ElGamal-Annapureddy-Veeravalli-arXiv12} for zero-forcing schemes. 

\begin{lem}
%[\protect{\cite[Lemma~3]{ElGamal-Annapureddy-Veeravalli-arXiv12}}]
 Consider any zero-forcing scheme. For any message $W_i$, the number of active receivers connected to at least one transmitter carrying the message is no greater than the number of transmitters carrying the message,
\begin{equation}\label{eq:zf1}
|{\cal V}_{{\cal T}_i}| \leq |{\cal T}_i|.
\end{equation}
Furthermore, the following has to hold.
\begin{equation}\label{eq:zf2}
\mathrm{rank}(D_{{\cal T}_{i},{\cal V}_{{\cal T}_i}}) = |{\cal V}_{{\cal T}_{i}}|.
\end{equation}
\label{lem:zf}
\end{lem}
\begin{IEEEproof}
We note that \eqref{eq:zf2} implies \eqref{eq:zf1}, but we include both in the theorem statement, and provide the proof of \eqref{eq:zf1} first for clarity. The statement of \eqref{eq:zf1} is the same as \cite[Lemma 3]{ElGamal-Annapureddy-Veeravalli-arXiv12}, but we briefly explain the proof here for completeness. Since we impose the constraint $I(W_i;Y_j)=0, \forall j\in{\cal V}_{{\cal T}_i}$, the interference seen at all receivers in ${\cal V}_{{\cal T}_i}$ has to be canceled. Also, since the probability of a zero Lebesgue measure set of channel realizations is zero, the $|{\cal T}_i|$ transmit signals carrying $W_i$ cannot be designed to cancel $W_i$ at more than $|{\cal T}_i|-1$ receivers for almost all channel realizations. This implies \eqref{eq:zf1}.

Now, we note that \eqref{eq:zf2} is equivalent to saying that there exists a matching between transmitters carrying $W_i$ and active receivers connected to transmitters carrying $W_i$, and this matching \emph{covers} all such active receivers. If this is not true while \eqref{eq:zf1} is satisfied, then it follows from Hall's Marriage Theorem~\cite{Hall} that there has to be subsets $\tilde{\cal T} \subset {\cal T}_i$, $\tilde{\cal V} \subset {\cal V}_{{\cal T}_i}$ such that $|\tilde{\cal T}|<|\tilde{\cal V}|$ and any transmitter whose index is in ${\cal T}_i \backslash \tilde{\cal T}$ is not connected to any receiver in $\tilde{\cal V}$. Hence, the above argument that we used to reach \eqref{eq:zf1} would apply if we consider the sets $\tilde{\cal T}$ and $\tilde{\cal V}$ as the set of transmitters carrying $W_i$ and the set of active receivers connected to them, respectively. It hence follows that \eqref{eq:zf2} holds, and the proof is thus complete.
\end{IEEEproof}
We now characterize the per user DoF for any zero-forcing scheme. 

\begin{thm}\label{thm:opt}
For any $K$-user hexagonal cellular network, the maximum achievable zero-forcing DoF under an average backhaul load constraint, $\eta_c^{\text{avg,zf}}(K,B)$ is the solution to the following optimization problem
\begin{align}
 \underset{\{{\cal T}_{j}\},\{d_{ij}\}_{i,j \in [K]}}{\text{max}}
 &\sum_{i\in[K]}\sum_{j\in[K]}  d_{ij}\\
 \text{s.t. }
 & d_{ij} \in \{0,1\},  \forall i,j\in [K],\label{con:d} \\
& d_{ij}=0, \text{if } i\notin {\cal N}_j \text{ or }  j\notin {\cal N}_i , \forall i,j\in [K],\label{con:d1}\\
&  \sum_{k\in {\cal N}_j} d_{kj} \leq1, \forall j\in [K],\label{con:d2}\\
& \sum_{k\in {\cal N}_j} d_{jk}\leq1,  \forall j\in [K],\label{con:d3}\\
&  d_{ij} \leq \mathbbm{1}\{i \in {\cal T}_{j}\},  \forall i,j\in [K],  \label{con:deliver} \\
& \frac{1}{K}\sum_j|{\cal T}_{j}|\leq B, \label{con:B} \\
%&\textcolor{Brown}{ |\tilde{\cal V}_{{\cal T}_{j}}| \leq |{\cal T}_{j}|},  \forall j\in [K], \label{con:lemma}\\
& \mathrm{rank}(D_{{\cal T}_{j},\tilde{\cal V}_{{\cal T}_j}}) = |\tilde{\cal V}_{{\cal T}_{j}}|,  \forall j\in [K], \label{con:mat} 
\end{align} 
where for any set ${\cal S} \subseteq [K]$, $\tilde{\cal V}_{\cal S}= \{k: k \in {\cal N}_i, i\in {\cal S} \text{ and } \mathbbm{1}\{\sum_{w\in {\cal N}_k} d_{wk}=1\}\}$.

%where any achievable zero-forcing scheme with $\{{\cal T}_{j}\},\{d_{ij}\}_{i,j \in [K]}$ is characterized by the constraints \eqref{con:d} - \eqref{con:mat} under the average backhaul load constraint $B$ with the per user DoF given by $\sum_{i\in[K]}\sum_{j\in[K]}  d_{ij}$.
\end{thm}
\begin{IEEEproof}
%We first note that for any achievable zero-forcing scheme $\sum_{i\in[K]}\sum_{j\in[K]}  d_{ij}$ gives the per user DoF. It suffices to show that any achievable zero-forcing scheme with average backhaul load $B$ is characterized by the constraints \eqref{con:d} - \eqref{con:mat}.

We first show that if the constraints in \eqref{con:d} - \eqref{con:mat} are satisfied, then there exists a message assignment satisfying the average backhaul load constraint $B$, and a zero-forcing scheme based on this assignment that achieves a per user DoF of $\sum_{i\in[K]}\sum_{j\in[K]}  d_{ij}$. It would follow then that $\eta_c^{\text{avg,zf}}(K,B) \geq \sum_{i\in[K]}\sum_{j\in[K]}  d_{ij}$, and hence the direct part of the theorem would be proved. 
It follows from \eqref{con:B} that the sets $\{{\cal T}_{j}\}_{j\in [K]}$ are transmit sets satisfying the average backhaul load constraint. We now construct the zero-forcing scheme. If $d_{ij}=1$, then we know from \eqref{con:deliver} that $i \in {\cal T}_j$ and we also know from \eqref{con:d1} that transmitter $j$ is connected to receiver $i$. We hence construct the transmit signal $X_{j,i}$ according to an optimal point-to-point code over an AWGN channel (see e.g.,~\cite{Cover-Thomas}) to deliver $W_i$ to its destination. We know from \eqref{con:d2} that $X_j$ would not be used to deliver any other message than $W_i$. Hence, we only need to show that interference caused by any such message $W_i$ at any active receiver can be canceled. From \eqref{con:mat}, we know that there is a matching between transmitters with indices in ${\cal T}_i$ and receivers with indices in ${\cal V}_{{\cal T}_i}$ that covers all such receivers. We hence assign a unique transmitter with an index $t\in{\cal T}_i \backslash \{j\}$ to each receiver with an index  $r\in{\cal V}_{{\cal T}_i} \backslash \{i\}$, and design the transmit signal $X_{t,i}$ to cancel the interference of $W_i$ at $Y_r$. Finally, it follows from \eqref{con:d3} that transmitter $j$ is the only transmitter connected to receiver $i$, and used to deliver $W_i$. It follows that we can achieve one degree of freedom for each binary variable $d_{ij}$, and hence, $\eta_c^{\text{avg,zf}}(K,B)$ is lower bounded by the solution of the optimization problem in the theorem statement. 
% For any achievable zero-forcing scheme, we have a message assignment $\{{\cal T}_{j}\}_{j \in [K]}$ with message $W_j$ sent interference free from some transmitter $i$ to receiver $j$ for some $j\in [K]$. 
 %Let $d_{ij}$ denote whether transmitter $i$ delivers the message to receiver $j$ interference-free i.e., \[ d_{ij}=\mathbbm{1}\{\text{Transmitter } i \text{ delivers message to receiver } j \text{ interference-free}\}.\] 
%Although message $W_j$ is available at $|{\cal T}_j|$ transmitters, only one transmitter delivers the message and the other transmitters use the message to zero-force interference at the receivers. Using two transmitters to deliver the same message does not increase the DoF achieved in the system, so we assume that each message is delivered by a single transmitter i.e., we have constraint \eqref{con:d2} or $d_{ij}\leq 1, \forall i,j\in [K]$.  Constraint \eqref{con:d} follows from the definition of $d_{ij}$ and constraint  \eqref{con:d1} follows because in order to deliver a message, a transmitter has to be connected to the receiver in the connectivity graph. Constraint \eqref{con:B} gives the backhaul load constraint. Constraint \eqref{con:deliver} ensures that a message cannot be delivered unless it is assigned and constraint \eqref{con:lemma} captures the statement of Lemma \ref{lem:zf}. 

We now describe the converse proof. Consider the optimal zero-forcing scheme achieving $\eta_c^{\text{avg,zf}}(K,B)$. We show that there is a choice of $\{{\cal T}_{j}\},\{d_{ij}\}_{i,j \in [K]}$ satisfying \eqref{con:d}-\eqref{con:mat} such that $\sum_{i\in[K]}\sum_{j\in[K]}  d_{ij} \geq \eta_c^{\text{avg,zf}}(K,B)$. Since the considered zero-forcing scheme satisfies the average backhaul load constraint of $B$, then \eqref{con:B} follows by setting $\{{\cal T}_{j}\}$ to the be the set of transmit sets of the considered scheme. Since we achieve zero degrees of freedom for every message whose receiver is inactive, the number of active receivers is at least the achieved degrees of freedom. We further know that since the definition of zero-forcing schemes in Section~\ref{sec:zf} ensures the creation of a point-to-point interference-free communication link for each active receiver, then there has to be an optimal zero-forcing scheme achieving $\eta_c^{\text{avg,zf}}(K,B)$, where we achieve one degree of freedom for each message corresponding to an active receiver; we assume that the considered scheme satisfies this property. For each active receiver with an index $i$, we can hence \emph{assign} a unique active transmitter with an index $j \in {\cal T}_i \cap {\cal N}_i$, such that $I(W_i;X_{j,i})>0$. If transmitter $i$ is assigned to receiver $j$, then we set $d_{ij}=1$. Otherwise, we set $d_{ij}=0$. We then have that \eqref{con:d}-\eqref{con:deliver} directly follow. 
Further, it follows that for any set ${\cal S} \subseteq [K]$, $\tilde{\cal V}_{\cal S}={\cal V}_{\cal S}$. We then have that \eqref{con:mat} follows from Lemma~\ref{lem:zf}, and the converse proof is thus complete.

\end{IEEEproof}

The optimization problem in Theorem~\ref{thm:opt} is difficult to solve numerically, because we are interested in the asymptotic behavior with large $K$, and the optimization is over a large number of message assignments, without an explicit bound on the maximum transmit set size constraint $M$. 
%For a fixed message assignment, we need to check the rank of a matrix for constraint \eqref{con:mat} for each message, which is of polynomial order in the size of the matrix and hence polynomial in $K$. Thus, for a fixed message assignment, verifying the constraints is polynomial in $K$. Now we try to find the number of message assignments. 
If a message assigned to $n$ transmitters where $0\leq n\leq K$, then we have $K \choose n$ possibilities to choose the transmit set, which is of the order ${\cal O}({\min(K^n,K^{K-n})})$. Since we consider a constraint on the average backhaul load and not the maximum transmit size, $n$ can be ${\cal O}(K)$ for a particular message. Thus, the computational complexity needed to just consider all message assignments is ${\cal O}(K^{\frac{K}{2}})$, i.e., exponential in $K$.\footnote{If we restrict our attention to the irreducible message assignments defined in \cite[Section V-D]{ElGamal-Annapureddy-Veeravalli-arXiv12}, then the complexity can be further reduced from ${\cal O}\left(K^{\frac{K}{2}}\right)$ to ${\cal O}\left(c^{\frac{K}{2}}\right)$, where $c$ is a constant that depends on the number of transmitters connected to a single receiver.} 

Hence, instead of trying to solve the optimization problem numerically, we focus on finding upper and lower bounds on the per user DoF.

\subsubsection*{{\bf Interference avoidance}}

We now restrict ourselves to $M=1$, and the class of interference avoidance schemes, which is a special case of zero-forcing schemes when $M=1$, and characterize lower and upper bounds for the maximum achievable per user DoF.

\begin{thm}\label{thm:tdma} 
The following bounds hold under restriction to interference avoidance schemes for the asymptotic per user DoF of hexagonal cellular networks with no cooperation,
%\begin{equation}
%\frac{1}{3}\leq\tau_c^{\text{zf}}(M=1)\leq\frac{3}{7}.
%\end{equation}
\begin{equation}
\frac{1}{3}\leq\tau_c^{\text{zf}}(M=1)\leq\frac{2}{5}.
\end{equation}
 \end{thm}

\begin{IEEEproof} The proof is available in Section \ref{subsec:tdma}.
\end{IEEEproof}

%\subsubsection{Zero-forcing schemes}
\subsubsection*{\bf Zero-forcing lower bounds}

We now allow for cooperation in the network and show through the results in Theorem~\ref{thm:main} and Theorem \ref{thm:ach2}, how a smart choice for assigning messages to transmitters, aided by cooperative transmission, can achieve scalable DoF gains through a zero-forcing coding scheme. For the achievable scheme in Theorem~\ref{thm:main}, this is done by treating the hexagonal network as interfering locally connected linear networks with connectivity parameter $L=2$, while the scheme in Theorem \ref{thm:ach2} considers a division of the network that does not involve linear networks. We note that it follows from Theorem \ref{thm:main} that we can achieve a per user DoF of $\frac{1}{2}$ without requiring an extra load on the backhaul ($B=1$), which is greater than the $\frac{2}{5}$ upper bound in Theorem \ref{thm:tdma} for the case without cooperation. 

%\textcolor{red}{
%\begin{thm}\label{thm:main}
%Under an integer backhaul load constraint $B$, the following lower bound holds for the asymptotic per user DoF of the hexagonal cellular network using zero-forcing schemes,
%\begin{equation}
% \tau_c^{\textup{avg,zf}}(B)\geq\frac{2B}{3B+1}, \forall B\in{\bf Z}^+.
%\end{equation}
%\end{thm}}
%\textcolor{red}{
%\begin{IEEEproof} 
%
%\end{IEEEproof}}

%
%\begin{thm}\label{thm:main}
%Under the average backhaul load constraint $B$, where $\frac{(5(\ell-1)+6)^2}{6\ell+9} < B \leq \frac{(5\ell+6)^2}{6\ell+9}$, for some $\ell\in\mathbb{N}\cup\{0\}$, the following lower bound holds for the asymptotic per user DoF of the hexagonal cellular network using zero-forcing schemes,
%\begin{equation*}
%\tau^{\text{avg,zf}}(B)\geq \text{max}\left\{ \sqrt{\frac{B}{6\ell+9}},\frac{2B}{3B+1}\right\}.
%\end{equation*}
%\end{thm}

%\begin{thm}\label{thm:main}
%Under the average backhaul load constraint $B$, where $\frac{(5\ell+6)^2}{6\ell+9} \leq B < \frac{(5(\ell+1)+6)^2}{6(\ell+1)+9}$, for some $\ell\in\mathbb{N}\cup\{0\}$, the following lower bound holds for the asymptotic per user DoF of the hexagonal cellular network using zero-forcing schemes,
%\begin{equation*}
%\tau^{\text{avg,zf}}(B)\geq \text{max}\left\{ \frac{5\ell +6}{6\ell+9},\frac{2B}{3B+1}\right\}.
%\end{equation*}
%\end{thm}

\begin{thm}\label{thm:main}
Under an integer backhaul load constraint $B$, the following lower bound holds for the asymptotic per user DoF of the hexagonal cellular network using zero-forcing schemes,
\begin{equation}
 \tau_c^{\textup{avg,zf}}(B)\geq\frac{2B}{3B+1}, \forall B\in{\bf Z}^+.
\end{equation}
\end{thm}

\begin{figure}[htb]
\centering
\includegraphics[scale=0.5]{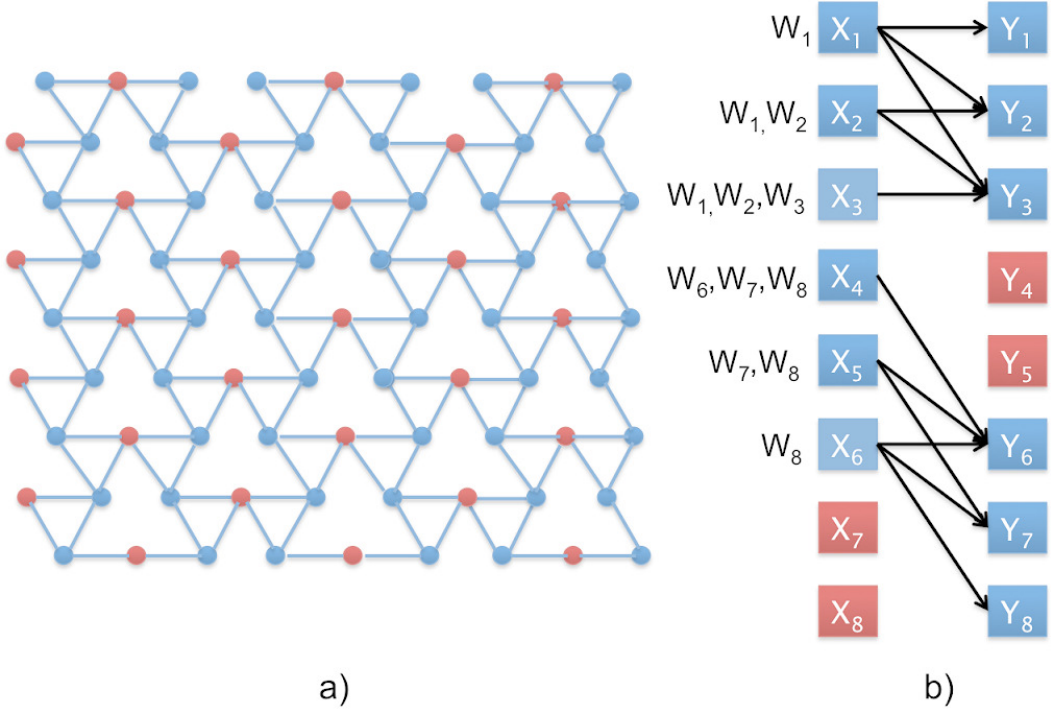} 
\protect\protect\caption{(a) Division of cellular network into subnetworks by deactivating nodes in $\Omega_{sq}$, and (b) the message assignment in each subnetwork for $B=1$. The unshaded nodes in (a) and the transmitters and receivers in the dashed boxes in (b) indicate that they are inactive.}
\label{fig_DivisionWyner}
\end{figure}

\begin{IEEEproof}

%The first part of the theorem follows from Lemma \ref{zf:achB} in Section \ref{subsec:main}. 

Consider a division of the network formed by deactivating the nodes in the set $\Omega_{sq}$ as shown in Figure \ref{fig_DivisionWyner}a. We note that the remaining network consists of non-interfering locally connected subnetworks with connectivity parameter $L=2$. 
In each subnetwork, we use the scheme in \cite{ElGamal-Annapureddy-Veeravalli-arXiv12} for $M=3B$ that considers a division of the subnetwork into non-interfering blocks of $6B+2$ nodes. The message assignment is shown in Figure \ref{fig_DivisionWyner}b for $B=1$. This scheme achieves a per user DoF of $\frac{M}{(M+1)}$ with $B=\frac{M}{2}$ in the locally connected linear subnetwork. Since the linear subnetworks only account for $\frac{2}{3}$ of the network, we obtain a per user DoF of $\frac{2B}{3B+1}$ with $B=\frac{M}{3}$ in the entire network.

\end{IEEEproof}

In Theorem \ref{thm:main},  $ \tau_c^{\textup{avg,zf}}(B) \to \frac{2}{3}$ as $B\to \infty$. We now consider achievable schemes which use a different division of the network and show that a per user DoF equal to $\frac{2}{3}$ can be achieved with $B=4$ with $ \tau_c^{\textup{avg,zf}}(B) \to \frac{5}{6}$ as $B\to \infty$.

\begin{thm}\label{thm:ach2}
Under the average backhaul load constraint $B$, where $\frac{(5\ell+6)^2}{6\ell+9} \leq B < \frac{(5(\ell+1)+6)^2}{6(\ell+1)+9}$, for some $\ell\in\mathbb{N}\cup\{0\}$, the following lower bound holds for the asymptotic per user DoF of the hexagonal cellular network using zero-forcing schemes,
\begin{equation}
\tau^{\text{avg,zf}}(B)\geq \frac{5\ell +6}{6\ell+9}.
\end{equation}
\end{thm}

\begin{IEEEproof}
 The proof is available in Section \ref{subsec:main}.
\end{IEEEproof}

\begin{figure}[htb]
\centering
\includegraphics[scale=0.45]{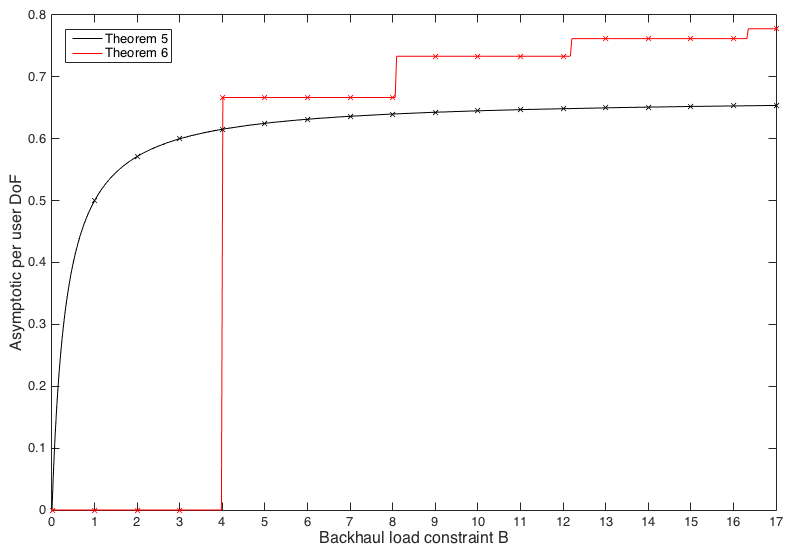} 
\protect\protect\caption{ Comparison of the lower bounds on the asymptotic per user DoF $\tau_c^{\textup{avg,zf}}(B)$.}
\label{fig:lbound}
\end{figure}

The achievable values for the per user DoF in Theorems \ref{thm:main} and \ref{thm:ach2} are compared in Figure \ref{fig:lbound}.

%We note that better DoF gains can be achieved for specific values of the average backhaul load $B$ using a linear combination of appropriate schemes for the average transmit set size constraint $M$ from Lemma \ref{zf:achB}.

\subsubsection*{\bf Zero-forcing upper bound}

%In this section, we discuss the problem of finding a zero-forcing upper bound under the average backhaul load constraint $B$. We first provide a lemma from \cite{ElGamal-Annapureddy-Veeravalli-arXiv12} for zero-forcing schemes and present the proof for Theorem \ref{zf:B}. We then formulate the problem of finding the maximum per user DoF under $B$ with restriction to zero-forcing schemes as an optimization problem.

 Let ${\cal K}_{in}$ denote the set of \emph{internal} nodes i.e., nodes which have five neighbors each, and ${\cal K}_{ex}$ denote the set of \emph{external} nodes i.e., nodes which have less than five neighbors.

\begin{figure}[htb]
\centering
\includegraphics[scale=0.37]{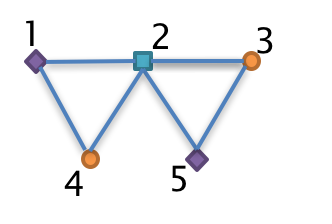} 
\protect\protect\caption{The figure illustrates the neighboring set ${\cal N}_j$ for $j=2$. }
\label{fig:block5}
\end{figure}

We now present the following upper bound on the per user DoF under the average backhaul load constraint $B$.

\begin{thm}\label{zf:B}
Under the average backhaul load constraint $B$, where $B < 5$, the following upper bound holds for the asymptotic per user DoF under restriction to zero-forcing schemes,
\begin{equation}
 \tau_c^{\textup{avg,zf}}(B)\leq\frac{1}{2}+\frac{B}{10}.
\end{equation}
\end{thm}
\begin{IEEEproof}

Consider any message assignment satisfying the average backhaul load constraint of $B$, and a zero-forcing scheme. Let  $\rho_j$ denote the fraction of users whose messages are available at exactly $j$ transmitters, where $0\leq j\leq K$. We have $\sum_{i=0}^{K} \rho_i=1$, and from the backhaul load constraint $B$, we have $\sum_{i=1}^{K}i \rho_i\leq B$. This gives us 
\begin{equation}\label{eq:ineqone}
\sum_{i=2}^{K}(i-1) \rho_i\leq (B-1)+\rho_0.
\end{equation}
We also note that for any given message assignment, the per user DoF is upper bounded by $1-\rho_0$.

As discussed in the proof of Theorem \ref{thm:opt}, it follows from our definition of zero-forcing schemes, that there is an optimal zero-forcing scheme where we achieve one degree of freedom for each message corresponding to an active receiver. Hence, for each active receiver with an index $i$, we can assign a unique active transmitter with an index $j \in {\cal T}_i \cap {\cal N}_i$, such that $I(W_i;X_{j,i})>0$.

Consider an active transmitter $j$ that is uniquely assigned to an active receiver $i$ such that $|{\cal T}_i|=m$ for some $1\leq m\leq 4$. In the set ${\cal N}_j$ (shown in Figure \ref{fig:block5}), where ${\cal N}_j$ denotes the set of five nodes adjacent to node $j$ including node $j$,
%In this block of five nodes, we have $\sum_{k\in {\cal N}_i} \textcolor{Brown}{r}_k\leq m$ for Lemma \ref{lem:zf} to hold.
from Lemma \ref{lem:zf} we have
\begin{equation}\label{eq:m4}\sum_{k\in {\cal N}_j} r_k\leq m, \quad 1\leq m\leq 4.\end{equation} 

Note that for any transmitter, the number of receivers in the neighboring set is five, and hence the number of active receivers is trivially upper bounded by five.
By summing the number of active receivers $\left(\sum_{k\in {\cal N}_j} r_k\right)$ in the neighboring set ${\cal N}_j$ over all the transmitters $j \in [K]$, we obtain the following. 
\begin{eqnarray}
\frac{5\sum_{i\in{\cal K}_{in}} r_{i}+c\sum_{i\in{\cal K}_{ex}}r_{i} }{K}& \overset{(a)}{\leq} & \sum_{i=1}^{4}i \rho_{i}+\sum_{i=5}^{K}5\rho_{i}+5\rho_{0}\\
 & \overset{(b)}{=} & 1+\sum_{i=2}^{4}(i-1)\rho_{i}+\sum_{i=5}^{K}4\rho_{i}+4\rho_{0}\\
 & \overset{(c)}{\leq} & 1+(B-1)+\rho_{0}+4\rho_{0},
\end{eqnarray}
where $c < 5$ is a positive constant, (a) follows from \eqref{eq:m4}, and $(b)$ follows since $\sum_{i=0}^{K}\rho_i=1$. Finally, $(c)$ follows because \eqref{eq:ineqone} implies that $\sum_{i=2}^{4}(i-1)\rho_i+\sum_{i=5}^{K}4\rho_i\leq (B-1)+\rho_0$. Note that each node in the interior of the graph has five neighbors, and hence appears five times on the left hand side of the inequality $(a)$.

We have $|{\cal K}_{ex}| ={\cal O}(\sqrt{K})$ which gives us $\frac{\sum_{j\in{\cal K}_{ex}}r_{j}}{K} =\frac{{\cal O}(\sqrt{K})}{K}\to 0$ as  $K\to \infty$. Thus, we have $\tau_c^{\textup{avg,zf}}(B)=\text{lim}_{K\to \infty}\frac{\sum_{i\in{\cal K}_{in}} r_i}{K} \leq \frac{B+5\rho_0}{5}$. It follows that for any message assignment, the per user DoF is upper bounded by min $\{1-\rho_0,\frac{B}{5}+\rho_0\}$ which gives us $ \tau_c^{\textup{avg,zf}}(B)\leq\frac{1}{2}+\frac{B}{10}$.

\end{IEEEproof}

\begin{figure}[htb]
\centering
\includegraphics[scale=0.45]{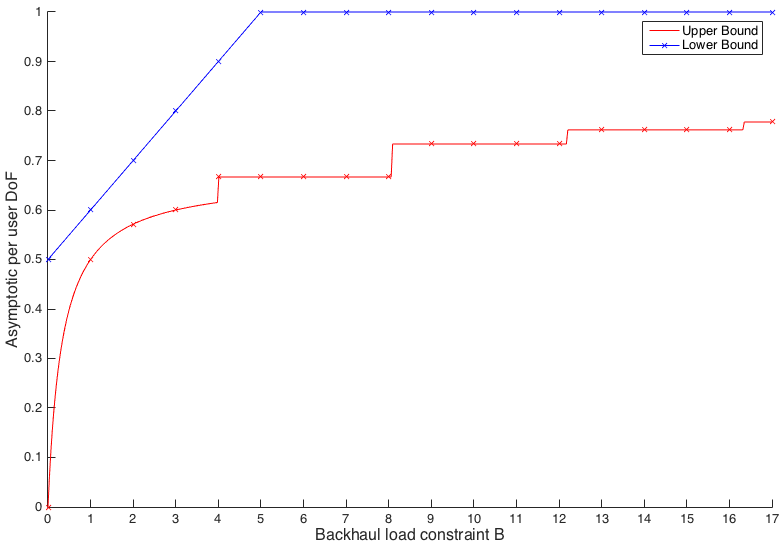} 
\protect\protect\caption{ Comparison of upper and lower bounds on the asymptotic per user DoF $\tau_c^{\textup{avg,zf}}(B)$.}
\label{fig:bound}
\end{figure}

We note that the bound in Theorem \ref{zf:B} may not be tight and is useful only for $B < 5$. The comparison between the upper and lower bounds for the per user DoF under zero-forcing schemes is shown in Figure \ref{fig:bound}. We believe that finding a general tight upper bound is difficult, especially for higher values of $B$, due to the combinatorial search for optimal message assignments as well as the rather complex connectivity structure of hexagonal networks. 

%In order to obtain solid insights into how a tight bound on the per user DoF can be obtained, we pose the problem of finding the maximum per user DoF under restriction to zero-forcing schemes as an optimization problem in the following section.

\subsection{Proof of Theorem \ref{thm:upper bound}}\label{subsec:upper bound}
In this section, before presenting the proof of Theorem \ref{thm:upper bound}, we first provide a lemma from \cite{ElGamal-Annapureddy-Veeravalli-arXiv12} for the case of $M=1$ that serves as a building block for the proof of Theorem \ref{thm:upper bound}, and then present a toy network for which we show that the per user DoF is upper bounded by $\frac{1}{2}$ in order to gain some insight into the proof of $\tau_c(M=1)\leq \frac{1}{2}$.

We present the following lemma for the case of $M=1$ which gives a relation between the DoF of the message being delivered by a transmitter and the DoF corresponding to the messages of the users connected to that transmitter. Here, ${\cal R}_j$ denotes the set of receivers that are connected to transmitter $j$. The lemma is an extension of the result in \cite{Host05} which shows that the maximum DoF for a
network with two transmitter-receiver pairs is unity. 

\begin{lem}
[\protect{\cite[Lemma~5]{ElGamal-Annapureddy-Veeravalli-arXiv12}}]
If ${\cal T}_{i}=\{X_{j}\}$, then $d_{i}+d_{k}\leq1$, $\forall k\in{\cal R}_{j}$.
\label{lem_m1}
\end{lem}

Each transmitter-receiver pair in the network is referred to as a node. If $a$ and $b$ are two nodes such that they are connected in the connectivity graph, and the transmitter of node $a$ has the message for node $b$, i.e., $a \in {\cal T}_b $, we denote this by $a\to b$.

\begin{figure}[htb]
\centering
\includegraphics[scale=0.5]{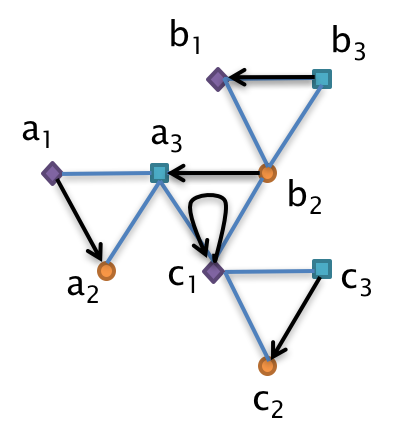}
\protect\caption{An example cellular network with nine transmitter-receiver pairs. The messages of $b_3,b_2,a_1,c_3$ can be assigned to any transmitter.}
\label{toy_example}
\end{figure}

\subsubsection*{Illustrative Example}

We consider the network and the message assignment shown in Figure \ref{toy_example} and show that the per user DoF in the network does not exceed $\frac{1}{2}$ for this particular message assignment. Note that the result holds for any assignment of the messages  $b_3,b_2,a_1,c_3$.
Since $a_1 \to a_2$, we have $d_{a_1}+d_{a_2} \leq 1$ from Lemma \ref{lem_m1}. Similarly, $b_3 \to b_1$ and $c_3 \to c_2$, we have $d_{b_1}+d_{b_3} \leq 1$ and $d_{c_2}+d_{c_3} \leq 1$, respectively from Lemma \ref{lem_m1}. We now show that $d_{a_3}+d_{b_2}+d_{c_1} \leq \frac{3}{2}$, and hence the per user DoF in this network is upper bounded by $\frac{1}{2}$. Note that ${\cal T}_{c_1}=\{c_1\}$ and hence $d_{b_2}+d_{c_1}\leq 1$ and $d_{a_3}+d_{c_1}\leq 1$. We also have $b_2 \to a_3$ and from Lemma \ref{lem_m1}, $d_{b_2}+d_{a_3}\leq 1$. Thus $d_{a_3}+d_{b_2}+d_{c_1} \leq \frac{3}{2}$ and the per user DoF in this network is upper bounded by $\frac{1}{2}$.

We now proceed with the proof of Theorem \ref{thm:upper bound}.

\begin{figure}[htb]
\centering
\includegraphics[scale=0.5]{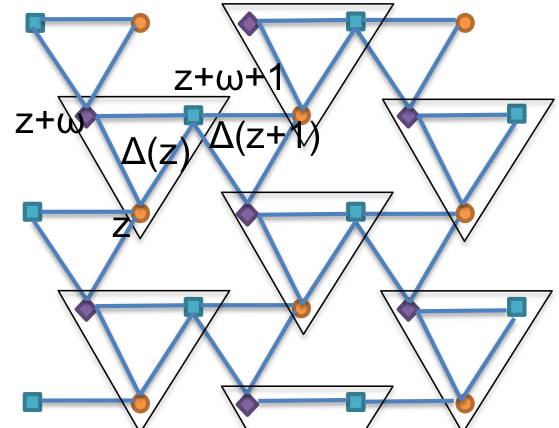}
\protect\caption{Division of the network into triangles.}
\label{proof_pic}
\end{figure}

%\begin{IEEEproof}   
Consider the division of the network into triangles ${\cal D}=\{\Delta(z) : z \in \Omega_{cir}\}$ as shown in Figure \ref{proof_pic}. 

For any $z \in \Omega_{cir}$, triangle $\Delta(z)$ consists of vertices $z,z+\omega,z+\omega+1$. Note that each triangle contains one vertex from each of the cosets, $\Omega_{cir}$, $\Omega_{sq}$ and $\Omega_{dia}$.

We refer to a node as a self serving node if the message to the receiver corresponding to the node is assigned to its own transmitter. 
We refer to a node as an outsider node if no message within its triangle is assigned to its transmitter, and also its message is not assigned within its triangle. Let $\cal O$ denote the set of outsider nodes given by,
\[{\cal O}=\{i\in \Delta(z): \Delta(z) \subseteq {\cal D}, {\cal T}_i \not \subseteq \Delta(z), {\cal T}_j\neq \{i\}, \forall j\in \Delta(z) \}.\] 
Without loss of generality, we assume that $|{\cal T}_j|=1, \forall j \in [K]$. Note that if the message of a particular receiver is not assigned to any transmitter, then the per user DoF cannot be increased if we assume that the message is assigned to any of the transmitters.
We say that a triangle is in state $S_i$ if exactly $i$ of the messages of the triangle are assigned to transmitters within the triangle, $0\leq i\leq3$. Let ${\cal S}_i$ denote the set of all triangles in state $S_i$. \[{\cal S}_i = \{ \Delta(z) \subseteq {\cal D}:  \mathbbm{1}_{\{ {\cal T}_z\subseteq \Delta(z)\}}+ \mathbbm{1}_{\{{\cal T}_{z+\omega}\subseteq \Delta(z)\}}+ \mathbbm{1}_{\{ {\cal T}_{z+\omega+1}\subseteq \Delta(z)\}}=i \}.\]
Let ${\cal SS}_1$ denote the set of all self serving nodes belonging to triangles in state ${S}_1$. More precisely,
\[{\cal SS}_1= \{ z: \Delta(z) \in {\cal S}_1, {\cal T}_z=\{z\}\} .\]

Note that every triangle in state $S_0$ consists of three outsider nodes, every triangle in state $S_1$ has at least one outsider node, and a triangle in state $S_2$ may contain an outsider node. 

We also define a middle triangle, as a triangle that is formed by the connected nodes of three different neighboring triangles. Middle triangles are triangles of the form $\{\Delta(z) : z \in \Omega_{dia}\}$. We say that a triangle is associated with a node if the node belongs to the triangle. If $z \in \Omega_{dia}$, the middle triangle associated with vertex $z$ is $\Delta(z)$. If $z \in \Omega_{sq}$, the middle triangle associated with vertex $z$ is $\Delta(z-\omega)$. If $z \in \Omega_{cir}$, the middle triangle associated with vertex $z$ is $\Delta(z-\omega-1)$. For any vertex $a$, we denote the middle triangle associated with it as $M_a$.
Note that each vertex is associated with exactly one main triangle and one middle triangle. We note that the definition of an outsider node is with respect to the main triangle associated with the node and not the middle triangle associated with it.

\begin{algorithm}[htb]
  \begin{algorithmic}[1]
  \caption{}
  \\Initialize ${\cal S} \gets \phi$ 
  \While {${\cal SS}_1\backslash {\cal S}\neq \phi$}
  \For{$a \in {\cal SS}_1$ where $a\in \Delta(z)$ for some $z \in \Omega_{cir}$}
 \State ${\cal S} \gets {\cal S}\cup\{a,j\}$ where $j = \underset {x \in \Delta(z)\backslash \{a\}}{\text {min}} \Re(x)$ 
  \EndFor
  \EndWhile
  \While {${\cal O}\backslash {\cal S} \neq \phi$}
  \For{$a \in {\cal O}\backslash {\cal S}$ where $a\in \Delta(z)$ for some $z \in \Omega_{cir}$ and the associated middle triangle $M_a$ contains nodes $b$ and $c$ apart from $a$.}
   \If {$M_{a}\backslash {\cal S}$ contains 3 outsider nodes}
    \State ${\cal S} \gets {\cal S}\cup\{a,b,c\}$
  \ElsIf{$M_{a}\backslash {\cal S}$ contains 2 outsider nodes $a$ and $j$ where $j\in \{b,c\}$}
    \State ${\cal S} \gets {\cal S}\cup\{a,j\}$
  \ElsIf{$M_{a}\backslash {\cal S}$ contains $a$ as the only outsider node and message for $a$ is assigned within $M_{a}\backslash {\cal S}$ at $j\in \{b,c\}$, i.e., $j \to a$}
    \State  ${\cal S} \gets {\cal S}\cup \{a,j\}$
  \ElsIf{$M_{a}\backslash {\cal S}$ contains $a$ as the only outsider node and message for $a$ is not assigned within $M_{a}\backslash {\cal S}$}
    \State ${\cal S} \gets {\cal S}\cup \{a\}$
  \EndIf  
  \EndFor
  \EndWhile
    \While {${\cal S}_1\cup {\cal S}_2 \cup {\cal S}_3\backslash S\neq \phi$}
  \For{triangle $T \in {\cal S}_1\cup {\cal S}_2 \cup {\cal S}_3$}
 \State ${\cal S} \gets {\cal S}\cup T\backslash {\cal S}$
  \EndFor
  \EndWhile
  \end{algorithmic}
\end{algorithm}

\begin{figure}[htb]
\centering
\includegraphics[scale=0.5]{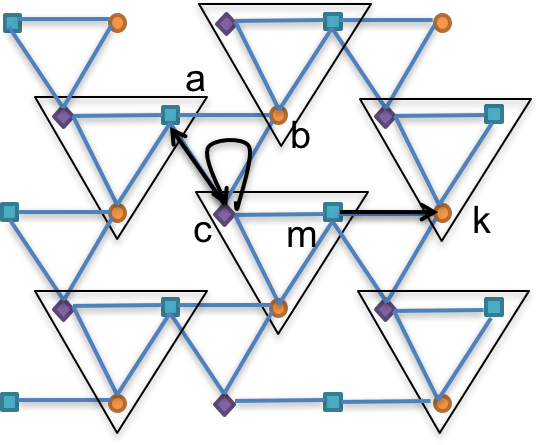}
\protect\caption{ We illustrate the case when $a$ is the only outsider node in its middle triangle and its message is available at $c$ where $c\in S$. The node $c$ is a self-serving node and node $m$ has been included according to line 4. The node $m$ contains the message for the only outsider node $k$ in the middle triangle containing $m$ and $k$.}
\label{proof_case1}
\end{figure}

Let $\tau_{\cal S}$ denote the per user DoF for the messages with indices in some set $\cal S$. We present Algorithm 1, to define a strategy for including nodes in a set ${\cal S}$, such that at any stage, the per user DoF of the nodes already included in ${\cal S}$ is upper bounded by $\frac{1}{2}$ i.e., $\tau_{\cal S}\leq \frac{1}{2}$. Note that at the end of the algorithm, all nodes are included in ${\cal S}$. To facilitate the understanding of Algorithm 1, we observe the following:
\begin{itemize}
\item If $a \in {\cal SS}_1$, then $a$ is a self-serving node and since the main triangle $T$ associated with it is in state $S_1$, the other nodes in the triangle $b,c$ are outsider nodes. We have $d_a+d_b \leq 1$ and $d_a+d_c \leq 1$, according to Lemma \ref{lem_m1}. Without loss of generality, we include the node with minimum real value among the two nodes $b,c$, and node $a$ in the set ${\cal S}$ as in line 4.

\item
%For any middle triangle with nodes $a,b,c$, containing at least two outsider nodes, from Lemma \ref{lem: mid} we have $d_a+d_b \leq 1$, $d_b+d_c \leq 1$, $d_a+d_c \leq 1$ and hence $d_a+d_b+d_c \leq \frac{3}{2}$. 
If $M_{a}\backslash {\cal S}$ where $M_a$ is a middle triangle, contains 3 outsider nodes, we include the nodes of that middle triangle $a,b,c$ in the set ${\cal S}$ as in line 10. If $M_{a}\backslash {\cal S}$ contains only two outsider nodes $a,j$, where $j\in \{b,c\}$, we include them in the set ${\cal S}$ as in line 12. 

We now show that if nodes are added to the set $\cal S$ according to line 10 or line 12, then the per user DoF of the nodes included in $\cal S$ is upper bounded by $\frac{1}{2}$.
In any middle triangle with nodes $a,b,c$, containing at least two outsider nodes, we show that $d_a+d_b \leq 1$, $d_b+d_c \leq 1$, $d_a+d_c \leq 1$ and hence $d_a+d_b+d_c \leq \frac{3}{2}$. Without loss of generality, let the two outsider nodes be $a$ and $b$. If the nodes $a,b$ are added according to line 12, it suffices to show that $d_a+d_b \leq 1$ whereas if the nodes $a,b,c$ are added according to line 10, we need to show that $d_a+d_b+d_c \leq \frac{3}{2}$.
 For node $a$, we have the following possibilities. \begin{itemize}
\item The message $W_a$ is not available at either $b$ or $c$. From our assumption, $W_a$ is not available at neighboring nodes outside the triangle. Hence, $W_a$ cannot be transmitted and we have $d_a=0$.
\item The message $W_a$ is available at one vertex in $b$ or $c$. From lemma \ref{lem_m1}, we have $d_a+d_c\leq 1$ and $d_a+d_b\leq 1$.
\end{itemize}
Similarly, for node $b$, we have $d_b=0$ if the message $W_b$ is not available at either $a$ or $c$, or $d_b+d_c\leq 1$ and $d_b+d_a\leq 1$ if the message $W_b$ is available at one vertex in $a$ or $c$. This gives us  $d_a+d_b+d_c\leq \frac{3}{2}$.

Thus, for any middle triangle with nodes $a,b,c$ with at least two outsider nodes, we have $d_a+d_b+d_c\leq \frac{3}{2}$. In addition, we also have $d_a+d_b \leq 1$, $d_b+d_c \leq 1$ and $d_a+d_c \leq 1$ as discussed above. Although for any middle triangle with at least two outsider nodes, the per user DoF is upper bounded by $\frac{1}{2}$, we do not include the third node in the set $\cal S$ in line 12 in order to simplify the cases considered later.

%In order to show this, let the two outsider nodes be $a$ and $T_1(cir)$. Since $T_1(cir)$ is an outsider node, we have the following possibilities. Suppose the message to node $T_1(cir)$ is not available at either $T_2(dia)$ or $T(i)$ and hence cannot be transmitted. Then we have $d_{T_1(cir)}=0$ and so $T(i)\circ T_1(cir)$ and $T_2(dia)\circ T_1(cir)$ hold. Suppose the message for $T_1(cir)$ is transmitted by any of the other nodes in the middle triangle. Then from lemma \ref{lem_m1}, we have $T(i)\circ T_1(cir)$ and $T_2(dia)\circ T_1(cir)$. Similarly, for the other outsider node $T(i)$, we have $T(i) \circ T_2(dia)$. Hence we have $a\circ T_1(cir) \circ T_2(dia)$.

\item
Let $a$ be the only outsider node in $M_{a}\backslash {\cal S}$, where $M_a$ is the middle triangle. If its message $W_a$ is available at neighboring node $j \in M_{a}\backslash {\cal S}$ where $j\in \{b,c\}$, i.e., $j \to a$, then we have $d_j +d_a \leq 1$ and include nodes $a,j$ in the set ${\cal S}$ as in line 14. 
\item
In the middle triangle $M_a$,
 if $W_a$ is not assigned within nodes $b,c$, we have $d_{a}=0$ and we include $a$ in the set ${\cal S}$ as in line 16.
%%%%%%%% this paragraph

We now consider the case where the message $W_a$ is assigned to a node in the set $M_{a}\cap {\cal S}$ and show that $\tau_{{\cal S}\cup\{a\}}\leq \frac{1}{2}$ when we add only the node $a$ in the set ${\cal S}$. Suppose $j\to a$ where $j \in \{b,c\}$ but $j\in {\cal S}$. We consider the case $j=c$ or $c \to a$  shown in Figure \ref{proof_case1}. So far, we have only added all outsider nodes in a few middle triangles and nodes from self-serving triangles. Hence this is possible only when $j$ was included in ${\cal S}$ according to line 4 in the algorithm. Without loss of generality, let $j$ be the self serving node and $m$ be the outsider node which was included in line 4. We have $d_j +d_a \leq 1$, $d_m +d_a \leq 1$ and we have $d_j +d_m \leq 1$ from before. Note that we have $d_m +d_a \leq 1$ from Theorem 1 since ${\cal T}_{a}=\{j\}$ and $m \in {\cal R}_j$. Hence $a$ can be included without any increase in the per user DoF. The same argument holds even if $j$ was the outsider node and $m$ the self serving node included in line 4. 
%Note that the only other possibility, through which nodes $j,j+1$ were previously included in the set ${\cal S}$ is when both $j$ and $j+1$ contain messages for the only remaining outsider nodes $T(sq)$ and $T'(k)$ in their respective middle triangles. In that case we see that $T(sq)\circ j$, $T'(k)\circ j+1$ and the per user DoF is still upper bounded by $\frac{1}{2}$.
Note that $j$ and $m$ could both contain messages for the only remaining outsider nodes $a$ and $k$ in their respective middle triangles. In that case we see that $d_j +d_a \leq 1$, $d_{k} +d_m \leq 1$ and  $\tau_{\cal S}\leq \frac{1}{2}$ when $k$ is added later according to line 16.
\end{itemize}

Consider all triangles in ${\cal S}_1\cup {\cal S}_2 \cup {\cal S}_3$. If $T$ denotes such a triangle with nodes $a,b,c$, let $t$ denote the set of nodes in $T$ but not included in ${\cal S}$ by line 19. For triangles in ${\cal S}_2, {\cal S}_3$ with nodes $a,b,c$, we have $d_a+d_b \leq 1, d_b+d_c \leq 1, d_a+d_c \leq 1$ and hence $d_a+d_b+d_c \leq \frac{3}{2}$ from Lemma \ref{lem_m1}. Consider the following cases for any triangle $T$ that has one or more nodes in the set $t=T \backslash {\cal S}$:
\begin{itemize}

\item
The set $t$ contains only one node $a$. We first find two nodes $b,j$ where $b\to j$ that were previously added to ${\cal S}$ according to line 14 and show that $d_j +d_a +d_b\leq \frac{3}{2}$ holds. We then show that nodes $b$ and $j$ do not appear in any other such combination, and hence $\tau_{\cal S}\leq \frac{1}{2}$ after adding $a$ to ${\cal S}$.

Note that by definition, a triangle in state $S_2$ or $S_3$ has at least two messages assigned within the triangle and thus has at least two non-outsider nodes. Hence, if $T \in {\cal S}_2\cup {\cal S}_3$, there exists at least one node  say $b$ such that $b$ is a non-outsider node and $d_a+d_b \leq 1$. We have the same conclusion if $T \in {\cal S}_1$, since all the self serving nodes and outsider nodes have already been included in ${\cal S}$. Hence, it is either the case that $a \to b$ or $b \to a$.

Since $b$ was a non-outsider node that was previously considered, it must have been added according to line 14. Hence, there is an assignment $b \to j$ where $j$ is an outsider node in the middle triangle $M_{b}$, $j\in \{a,c\}$ and $d_b+d_j \leq 1$ was considered. We also have $d_a+d_b \leq 1$ and $d_a +d_j\leq 1$ from Lemma \ref{lem_m1} since ${\cal T}_{j}=\{b\}$ and $a \in {\cal R}_{b}$. Hence we have $d_j +d_a +d_b\leq \frac{3}{2}$. 

Note that neither $j$ nor $b$ is part of any other such combination. This is true for $b$ because all the nodes in its triangle have already been considered. Since $b\to j$ and $j$ has been added to the set ${\cal S}$ according to line 14, outsider node $j$ cannot be part of any such combination that does not involve $b$. Thus, we include $t=\{a\}$ in the set ${\cal S}$ as in line 22 while maintaining  $\tau_{\cal S}\leq \frac{1}{2}$.

\item The set $t$ contains two nodes say $a,b$.
If $T\in {\cal S}_1$, then either $a\to b$ or $b\to a$ and we have $d_a+d_b \leq 1$. 
If $T \in {\cal S}_2\cup {\cal S}_3$, we have $d_a+d_b \leq 1$ and we include $t=\{a,b\}$ in the set $S$ as in line 22. 

\item The set $t$ contains three nodes $a,b,c$.
This can happen only when $T \in {\cal S}_2\cup {\cal S}_3$. In this case, we have $d_a+d_b+d_c \leq \frac{3}{2}$ and we include $t=\{a,b,c\}$ in the set ${\cal S}$ as in line 22. 
% and in this case, we have $<T(1),T(2),T(3)>$ and we include $t=\{T(1),T(2),T(3)\}$ in the set $S$ as in line 22. 
\end{itemize} 
%\end{IEEEproof}

\subsection{Proof of Theorem \ref{thm:tdma}}\label{subsec:tdma}
%\color{ForestGreen}
%Consider the following lemma from \cite{ElGamal-Annapureddy-Veeravalli-arXiv12}
%\begin{lem}\label{lem:zf}
% For any message $W_i$, the number of active receivers connected to at least one transmitter carrying the message is no greater than the number of transmitters carrying the message,
%\[|V_{{\cal T}_i}| \leq |{\cal T}_i|.\]
%\end{lem}

% In the network let ${\cal K}_{in}$ denote the set of interior nodes which have five neighbors each and ${\cal K}_{ex}$  denote the set of exterior nodes which have less than five neighbors.

\begin{figure}[htb]
\centering
\subfloat[]{\includegraphics[height=0.27\textwidth]{M1division}}                
\quad\quad\subfloat[]{\includegraphics[width=0.1\textwidth]{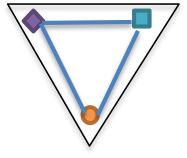}}
  \caption{Division of network into triangular subnetworks in (a). In (b), we note that by deactivating square and diamond nodes, a per user DoF of $\frac{1}{3}$ is achieved.}
\label{DivisionTri}
\end{figure}

%For each node $i$, let $\textcolor{Brown}{t}_i$, $\textcolor{Brown}{r}_i$ indicate whether transmitter $i$ and receiver $i$ are active or not i.e., 
%$\textcolor{Brown}{t}_{i}=\mathbbm{1}\{$Transmitter $i$ is active$\}$ and  $\textcolor{Brown}{r}_{i}=\mathbbm{1}\{$Receiver $i$ is active$\}$. 
%%If transmitter $i$ is active, then $\textcolor{Brown}{t}_i=1$, otherwise it is zero. Similarly for $\textcolor{Brown}{r}_i$. 
%We note that sumDoF = $ \frac{1}{2} \sum_i (\textcolor{Brown}{t}_i+\textcolor{Brown}{r}_i)$. 

 \paragraph*{Lower Bound} 
 
Consider the division of the network into triangles ${\cal D}=\{\Delta(z) : z \in \Omega_{cir}\}$ as shown in Figure \ref{DivisionTri}. For any $z \in \Omega_{cir}$, triangle $\Delta(z)$ consists of vertices $z,z+\omega,z+\omega+1$.
%Consider the division of the network into disjoint fully connected
%triangles as shown in Figure \ref{DivisionTri}}. 
By deactivating the nodes $\{z: z \in \Omega_{sqr}\cup \Omega_{dia}\}$, i.e., the square and diamond nodes in each triangle, the network decomposes into $\frac{K}{3}$ isolated nodes $\{z:z \in \Omega_{cir}\}$ that each achieves a DoF of one, thus achieving a per user DoF of $\frac{1}{3}$ in the network.

\paragraph*{Upper Bound}

For each node $j \in {\cal K}_{in}$ in the interior of the network, consider the set of neighbors ${\cal N}_j$. This results in the block of five nodes as shown in Figure \ref{fig:block5}. For any such $j$, we show that 
\[\sum_{i\in {\cal N}_j} \{r_i + t_i \}\leq 4.\]
 
For any zero-forcing scheme, we first note that among any two adjacent nodes $i,j$, 
\[r_i + t_i + \textcolor{Brown}{r}_j + t_j\leq 2,\] i.e., among any two adjacent nodes, at least two transmitters or receivers are inactive. This holds because if one of the receivers is active, one transmitter has to be inactive among the nodes $\{i,j\}$ and if one of the transmitters is active, one of the receivers among the nodes $\{i,j\}$ has to be inactive. 

We further note that
any fully connected triangle in the network is in one of the following
states:

State 0 (inactive triangle): All transmitters and receivers in the
triangle are inactive.

State 1 (self-serving triangle): Exactly one transmitter in the triangle
sends a message to exactly one receiver within the triangle. None of
the other transmitters or receivers can be active in this triangle.

State 2 (serving triangle): At least one transmitter in the triangle
is activated to serve a receiver in another triangle and there are
no active receivers in the considered triangle.

State 3 (served triangle): At least one receiver in the triangle is
activated as it is being served by a transmitter in another triangle
and there are no active transmitters within the considered triangle.

Without loss of generality we now consider $j=2$ and the block of five nodes shown in Figure \ref{fig:block5} and show that $\sum_{i\in {\cal N}_2} \{r_i + t_i \}\leq 4$. We show that at least six transmitters or receivers must be inactive. 
Consider the triangle formed by nodes $\{1,2,4\}$:
\begin{itemize}
\item If the triangle is in State 0 then all three transmitters and receivers are inactive and we are done. 

\item If the triangle is in State 1, then among the three nodes, there is at least one inactive node. Among the remaining adjacent nodes in the triangle at least two of the transmitters or receivers are inactive. Among the nodes $\{3,5\}$, at least two of the transmitters or receivers are inactive. Thus in the block of five nodes, $\sum_{i\in {\cal N}_2} \{r_i + t_i \}\leq 4$.

\item If the triangle is in State 2, then all three receivers in the triangle are inactive. Suppose all three transmitters in the triangle are active. Then one receiver among nodes $\{3,5\}$ must be receiving message from transmitter $2$ and the remaining node among $\{3,5\}$ is inactive. Thus at least six transmitters or receivers are inactive. If on the other hand, at least one transmitter in the triangle is inactive, then we have three inactive receivers and one inactive transmitter within the triangle. Among the nodes $\{3,5\}$, at least two of the transmitters or receivers are inactive. Thus in the block of five nodes, $\sum_{i\in {\cal N}_2} \{r_i + t_i \}\leq 4$.
 
\item If the triangle is in State 3, the discussion follows in a similar fashion to the State 2 case with transmitters instead of receivers.

\end{itemize}

Summing this up over all $K$ users, for some constant $c< 5$, we have \[5\sum_{i\in {\cal K}_{in}} \{r_i + t_i \}+ c \sum_{j\in {\cal K}_{ex} }\{r_j+ t_j\} \leq 4K.\]  %We have that $|{\cal K}_{ex}| ={\cal O}(\sqrt{K})$ which gives us per user DoF less than or equal to $\frac{2}{5}$. 
We have $|{\cal K}_{ex}| ={\cal O}(\sqrt{K})$ which gives us \[\frac{\sum_{j\in{\cal K}_{ex}}\{r_{j}+ t_j\}}{K} =\frac{{\cal O}(\sqrt{K})}{K}\to 0 \text{ as }  K\to \infty.\] Thus, we have \[\tau_c^{\textup{avg,zf}}(M=1)\leq \text{lim}_{K\to \infty}\frac{\sum_{i\in{\cal K}_{in}} r_i+ t_i}{2K},\] which gives us per user DoF less than or equal to $\frac{2}{5}$.

\subsection{Proof of Theorem \ref{thm:ach2}}\label{subsec:main}

We first show that under the maximum transmit set size constraint $M$ defined in \eqref{M}, where $5(\ell-1)+6 < M \leq 5\ell+6$, for some $\ell\in\mathbb{N}\cup\{0\}$, a per user DoF of $\frac{M}{6\ell+9}$ can be achieved
with an average backhaul load $B=\frac{M^{2}}{6\ell+9}$ and the proof follows.

%\begin{lem}\label{zf:achB}
%Under the maximum transmit set size constraint $M$ defined in \eqref{M}, where $5(\ell-1)+6 < M \leq 5\ell+6$, for some $\ell\in\mathbb{N}\cup\{0\}$, the following lower bound holds on the per user DoF, as the number of users goes to infinity,
%\begin{equation*}
%\tau^{\text{zf}}_c(M)\geq \frac{M}{6\ell+9},
%\end{equation*}
%\end{lem}
%with an average backhaul load $B=\frac{M^{2}}{6\ell+9}$.

For any $\ell$, consider the division of the network into blocks of $6\ell+9 $ nodes by deactivating the nodes in the set ${\cal D}_{\ell}$, defined as,
$${\cal D}_{\ell}=\left\{\Delta(z)\bigcup_{m\in[\ell]}\{z-\sqrt{3}m\imath\} : z \in {\cal G} \right\},$$ 
where
%${\cal G}=\{z: \text{Im}(z)= (2\ell+1)k, \text{Re}(z)= \frac{3}{2}k+p, \forall k, p\in \mathbbm{Z} \}.$
%${\cal G}=\{z: z = (\frac{3}{2}k+p)+i((2\ell+1)k), \forall k, p\in \mathbbm{Z} \}$, where $i = \sqrt{-1}$.
${\cal G}=\{z: z = (\frac{3}{2}k+3p)+\imath((2\ell+3)\frac{\sqrt{3}}{2}k), \forall k, p\in \mathbbm{Z} \}$, where $\imath = \sqrt{-1}$.

We first prove the result for $1<M\leq 6$ for which $\ell=0$ and then extend
this scheme to higher values of $M$ which correspond to higher values of $\ell$. 
%Consider the division of the network into blocks of nine nodes by deactivating the nodes ${\cal D}=\{\Delta(z) : z \in \Omega_{sq}, \text{Im}(z)= 3k\frac{\sqrt{3}}{2}, \text{Re}(z)= 3m, k,m\in \mathbbm{Z} \}$ as shown in Figure \ref{NetworkDivision}. 
By deactivating nodes in ${\cal D}_0$ the network decomposes into non-interfering blocks containing six nodes each.
In the block of six nodes, if $M$ messages are each available at $M$ 
transmitters, then by the use of simple linear transmit beamforming, we 
obtain a sum DoF of $M$ thus giving us a per user DoF of $\frac{M}{9}$. Note that for this scheme, the average backhaul load on the network $B=\frac{M^2}{9}$.

%\begin{figure}[htb]
%\centering 
%\includegraphics[scale=0.4]{Linear}
%\protect\protect\caption{The subnetwork is shown as a linear network and the message assignment in the linear network is shown for $M=2$. The red boxes indicate that the corresponding transmitter/receiver is deactivated.}
%\label{Linear}
%\end{figure}

\begin{figure}[htb]
\centering 
 \includegraphics[scale=0.50]{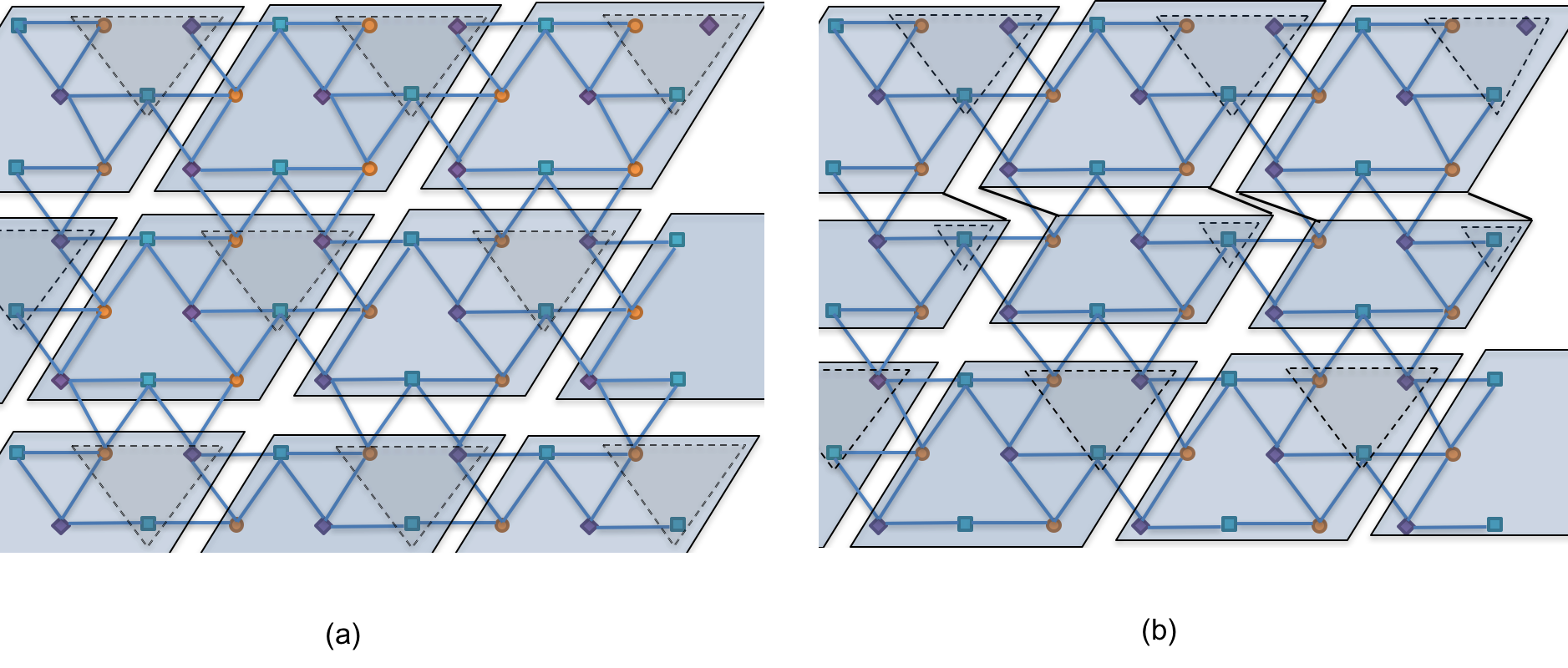} 
\protect\protect\caption{Division of cellular network into subnetworks. In (a), $\ell = 0$ and each block has six nodes each. In (b), $\ell=1$ and each block has a sub-block containing nine nodes and a sub-block containing six nodes below it. The nodes in the triangles denote the deactivated nodes in the network.}
\label{fig:ubl}
\end{figure}

%Consider the division of the network into blocks of nine nodes by deactivating the nodes ${\cal D}_{\ell}=\{z: \text{Re}(z)= 3m, m\in \mathbbm{Z} \}\cup\{\Delta(z) : \text{Im}(z)= \frac{\sqrt{3}}{2}2\ell k, \text{Re}(z)= 3m, m,k\in \mathbbm{Z}\}$ as shown in Figure \ref{}. 
For a higher $M$ such that $5(\ell-1)+6 < M \leq 5\ell+6$ with $\ell\geq$ 1, consider subnetworks of size $9+ 6\ell$.
The case $\ell=1$ is shown in the Figure \ref{fig:ubl}(b).
By deactivating the nodes in ${\cal D}_{\ell}$ the network decomposes into non-interfering blocks containing $5\ell+6$ nodes each. 
In each non-interfering block, 
we have a sub-block of six nodes as in the previous
case and $\ell$ sub-blocks containing five nodes each. If in each block,  $M$ messages are each available at $M$ 
transmitters, then by the use of simple linear transmit beamforming, we 
obtain a sum DoF of $M$ in each block of ${6\ell+9}$ nodes.
Thus a per user DoF of $\frac{M}{6\ell+9}$ can be attained 
with an average backhaul load of $\frac{M^{2}}{6\ell+9}$.

\section{Conclusion}\label{sec:conclusion}
We studied the potential gains offered by cooperative transmission in the downlink of cellular networks, under an average backhaul load constraint. We first characterized the asymptotic per user DoF in the linear interference network
and showed that the optimal coding scheme relies only on zero-forcing transmit beamforming. The optimal schemes rely on an asymmetric assignment of messages, such that the backhaul constraint is satisfied, where some messages are assigned to more than $B$ transmitters, others are assigned to fewer than $B$ transmitters, and the remaining messages are not assigned at all. Thus, the average backhaul constraint allows for higher degree of freedom gains compared to the maximum transmit set size constraint and hence we have $\tau^{\text{avg}}(B)>\tau(M)$.
We then extended these results to the more general and practically relevant hexagonal sectored cellular networks. We showed that DoF gains can be achieved using cooperative transmission under the average backhaul load constraint $B$ by proposing achievable schemes for general integer values of $B$. 
The proposed schemes are simple zero-forcing schemes with a flexible message assignment that achieve the information-theoretic upper bound of the per user DoF when cooperation is not allowed. Further, in order to achieve this bound, there is neither need to increase the backhaul load beyond an average of one message per transmitter, nor to use interference alignment.
 %The proposed schemes are simple zero-forcing schemes with a flexible message assignment that achieve the information-theoretic upper bound of the per user DoF for the case of no-cooperation, with an average backhaul load of one message per transmitter, i.e., with no extra backhaul load \textcolor{red}{($\tau_c^{\text{avg,zf}}(B=1)\geq \tau_c(M=1)$)}, without the need for interference alignment.
We also showed that $\tau_c^{\text{zf}}(M=1)<\tau_c(M=1)$, i.e., interference avoidance schemes cannot achieve the information theoretic upper bound of $\frac{1}{2}$, a DoF value that \emph{can} be achieved with zero-forcing \em{cooperative } transmission and no extra backhaul load. Further, we provided a useful upper bound on the per user DoF achievable through cooperative zero-forcing with small values of the backhaul load $B<5$, $\tau_c^{\text{avg,zf}}(B)\leq \frac{5+B}{10}$. In order to obtain a tight bound on the per user DoF for zero-forcing schemes for any backhaul load constraint $B$, we formulated the general problem of finding the maximum per user
DoF as an optimization problem.

It is important to note that the conclusions in this work, rely on the assumption that accurate channel state information (CSIT) is available at the transmitters. Recently, the problem of interference management through cooperative transmission has been studied with weak and no CSIT in~\cite{Gesbert-weakcsit}-\cite{Jafar-solution}. In~\cite{Gesbert-nocsit}, it was shown that significant gains could be achieved through a flexible cell association strategy that does not constrain availability of the $i^{th}$ message to only the $i^{th}$ transmitter. In~\cite{spawc}, it was shown that cooperative transmission cannot lead to a per user DoF gain in large Wyner's linear networks with no CSIT, when restricted to linear cooperation schemes. It is of interest to extend the work in \cite{Gesbert-weakcsit}-\cite{Jafar-solution} to study interference management using cooperative transmission with weak and no CSIT. %The benefit of cooperation with lack of accurate CSIT is still questionable in light of recent research attempts.
% The proposed schemes are simple zero-forcing schemes with a flexible message assignment that achieve the information theoretic upper bound of the per user DoF for the case of no-cooperation with an average backhaul load of one message per transmitter, i.e., with no extra backhaul load \textcolor{red}{($\tau_c^{\text{avg,zf}}(B=1)\geq \tau_c(M=1)$)}, without the need for interference alignment.

\bibliographystyle{IEEEtran}

\section*{appendix}

\section*{Upper bound in Theorem 1}

%\subsubsection{Upper Bound}\label{sec:ub}
We prove the upper bound in Theorem~\ref{thm:main_result} in two steps. First, we provide an information theoretic argument in Lemma~\ref{lem:aux_ub} to prove an upper bound on the DoF of any network that has a subset of messages whose transmit set sizes are bounded. We then finalize the proof with a combinatorial argument that shows the existence of such a subset of messages in any assignment of messages satisfying the backhaul constraint of~(\ref{B}).

In order to establish the information theoretic argument in Lemma~\ref{lem:aux_ub}, we use Lemma~\ref{lem:dofouterbound}, that is introduced in Section~\ref{sec:proof}. We also need~\cite[Corollary $3$]{ElGamal-Annapureddy-Veeravalli-arXiv12} in the proof of Lemma~\ref{lem:aux_ub}; we restate it for the considered system model.
\begin{cor}
[\protect{\cite[Corollary 3]{ElGamal-Annapureddy-Veeravalli-arXiv12}}]
%[\cite{ElGamal-Annapureddy-Veeravalli-arXiv12}]
\label{cor:dofouterbound}
For any $K$-user linear interference channel, if the size of the transmit set $|{\cal T}_i| \leq M, i\in[K],$ then any element $k \in {\cal T}_i$ such that $k \notin \{i-M,i-M+1,\ldots,i+M-1\}$ can be removed from ${\cal T}_i$, without decreasing the sum rate.
\end{cor}

We now make the following definition to use in the proof of the following lemma. For any set ${\cal S}\subseteq [K]$, let $g_{\cal S}: {\cal S} \rightarrow \{1,2,\ldots,|{\cal S}|\}$ be a function that returns the ascending order of any element in the set ${\cal S}$, e.g., $g_{\cal S}\left(\min \left\{i: i\in{\cal S}\right\}\right)=1$ and $g_{\cal S}\left(\max \left\{i: i\in{\cal S}\right\}\right)=|{\cal S}|$

\begin{lem}\label{lem:aux_ub}
For any $K$-user linear interference channel with DoF $\eta$, if there exists a subset of messages ${\cal S} \subseteq [K]$ such that each message in ${\cal S}$ is available at a maximum of $M$ transmitters, i.e., $|{\cal T}_i| \leq M, \forall i \in {\cal S}$, then the DoF is bounded by,
\begin{equation}
\eta \leq K-\frac{|{\cal S}|}{2M+1}+C_K,
\end{equation}
where $\lim_{K \rightarrow \infty} \frac{C_K}{K} = 0$.
\end{lem}
\begin{IEEEproof}
We use Lemma~\ref{lem:dofouterbound} with a set ${\cal A}$ such that the size of the complement set $|\bar{\cal A}|=\frac{|{\cal S}|}{2M+1}-o(K)$. We define the set ${\cal A}$ such that ${\bar{\cal A}}=\{i: i\in{\cal S}, g_{\cal S}(i)= (2M+1)(j-1)+M+1, j \in {\bm{Z}^+}\}$. 

Now, we let $s_1$, $s_2$ be the smallest two indices in $\bar{\cal A}$. We see that $g_{\cal S}(s_1)=M+1, g_{\cal S}(s_2)=3M+2$. Note that $X_1+\frac{Z_1}{H_{1,1}}=\frac{Y_1}{H_{1,1}}$, and
\begin{equation*}
X_2+\frac{Z_2-\frac{H_{2,1}}{H_{1,1}}Z_1}{H_{2,2}}=\frac{Y_2-\frac{H_{2,1}}{H_{1,1}}Y_1}{H_{2,2}}.
\end{equation*}
Similarly, it is clear how the first $s_1-1$ transmit signals $X_{1},X_{2},\ldots,X_{s-1}$ denoted as $X_{[s_1-1]}$ can be recovered from the received signals $Y_{[s_1-1]}$ and linear combinations of the noise signals $Z_{[s_1-1]}$. In what follows, we show how to reconstruct a noisy version of the signals $\left\{X_{s_1},X_{s_1+1},\ldots,X_{s_2-1}\right\}$, where the reconstruction noise is a linear combination of the signals $Z_{\cal A}$. Then it will be clear by symmetry how the remaining transmit signals can be reconstructed. 

We now notice that it follows from Corollary~\ref{cor:dofouterbound} that message $W_{s_1}$ can be removed from any transmitter in ${\cal T}_{s_1}$ whose index is greater than $s_1+M-1$, without affecting the sum rate. Similarly, there is no loss of generality in assuming that $\forall s_i\in{\cal S}, s_i \neq s_1$, ${\cal T}_{s_i}$ does not have an element with index less than $s_i-M$. Since $s_i-s_1\geq g_{\cal S}(s_i)-g_{\cal S}(s_1)\geq 2M+1$, it follows that $X_{s_1+M} \in X_{U_{\cal A}}$. The signal $X_{s_1+M+1}+\frac{Z_{s_1+M+1}}{H_{s_1+M+1,s_1+M+1}}$ can be reconstructed from $Y_{s_1+M+1}$ and $X_{s_1+M}$. Then, it can be seen that the transmit signals $\left\{X_{s_1+M+2},X_{s_1+M+3},\ldots,X_{s_2-1}\right\}$ can be reconstructed from $\left\{Y_{s_1+M+1}, Y_{s_1+M+2},\ldots,Y_{s_2-1}\right\}$, and linear combinations of the noise signals $\left\{Z_{s_1+M+1},Z_{s_1+M+2},\ldots,Z_{s_2-1}\right\}$. Similarly, since $X_{s_1+M}$ is known, the transmit signals $\left\{X_{s_1+M-1},X_{s_1+M-2},\ldots,X_{s_1}\right\}$ can be reconstructed from $\left\{Y_{s_1+M}, Y_{s_1+M-1},\ldots,Y_{s_1+1}\right\}$, and linear combinations of the noise signals $\left\{Z_{s_1+M},Z_{s_1+M-1},\ldots,Z_{s_1+1}\right\}$. By following a similar argument to reconstruct all transmit signals from the signals $Y_{\cal A}$, $X_{U_{\cal A}}$, and linear combinations of the noise signals $Z_{\cal A}$, we can show the existence of functions $f_1$ and $f_2$ of Lemma~\ref{lem:dofouterbound} to complete the proof.
\end{IEEEproof}

We now explain how Lemma~\ref{lem:aux_ub} can be used to prove that $\tau^{\text{avg}}(B=1) \leq \frac{3}{4}$. For any message assignment satisfying~(\ref{B}) for a $K$-user channel, let $\rho_{j}$ be defined as follows for every $j\in\{0,1,\ldots,K\}$,
\begin{equation}\label{eq:concentration}
\rho_{j} = \frac{|\left\{i:i\in[K], |{\cal T}_i|=j\right\}|}{K}.
\end{equation}
$\rho_{j}$ is the fraction of users whose messages are available at exactly $j$ transmitters. Now, if $\rho_0+\rho_1 \geq \frac{3}{4}$, then Lemma~\ref{lem:aux_ub} can be used directly to show that $\eta \leq \frac{3K}{4}+o(K)$. Otherwise, more than $\frac{K}{4}$ users have their messages at two or more transmitters, and it follows from~(\ref{B}) that $\rho_0 \geq \sum_{j=2}^K \rho_j \geq \frac{1}{4}$, and hence, $\eta \leq (1-\rho_0)K \leq \frac{3K}{4}$.

We generalize the above argument to complete the proof that $\tau^{\text{avg}}(B)\leq\frac{4B-1}{4B},\forall B\in \bm{Z}^+$. More specifically, we show that
for any message assignment satisfying~(\ref{B}) for a $K$-user channel with an average transmit set size constraint $B$, there exists an integer $M\in\{0,1,\ldots,K\}$, and a subset ${\cal S} \subseteq [K]$ whose size $|{\cal S}|\geq \frac{2M+1}{4B} K$, such that each message in ${\cal S}$ is available at a maximum of $M$ transmitters, i.e., $|{\cal T}_i| \leq M, \forall i\in{\cal S}$.
Fix any message assignment satisfying~(\ref{B}) for a $K$-user channel with backhaul constraint $B$, and let $\rho_j,j\in\{0,1,\ldots,K\}$ be defined as in~\eqref{eq:concentration}. If $\sum_{j=2B}^{K} \rho_j \leq \frac{1}{4B}$, then more than $\frac{4B-1}{4B}K$ users have a transmit set whose size is at most $2B-1$, and the statement follows with $M=2B-1$. It is then possible to assume that $\sum_{j=2B}^{K} \rho_j > \frac{1}{4B}$. In what follows, we show by contradiction that there exists an integer $M \in\{0,\ldots,2B-2\}$ such that $\sum_{j=0}^M \rho_j > \frac{2M+1}{4B}$.

Define $\rho_j^*, j\in\{0,1,\ldots,2B\}$ such that $\rho_0^*=\rho_{2B}^*=\frac{1}{4B}$, and $\rho_j^*=\frac{1}{2B}, \forall j\in\{1,\ldots,2B-1\}$. Now, note that $\sum_{j=0}^{2B} \rho_j^*=1$, and $\sum_{j=0}^{2B} j\rho_j^*=B$. It follows that if $\rho_j=\rho_j^*, \forall j\in\{0,\ldots,2B\}$, and $\rho_j=0, \forall j \geq 2B+1$, then the constraint in~(\ref{B}) is tightly met, i.e., $\frac{\sum_{i=1}^K |{\cal T}_i|}{K}=B$. We will use this fact in the rest of the proof. 

Assume that $\sum_{j=2B}^K \rho_j > \rho_{2B}^* = \frac{1}{4B}$, and that $\forall M\in\{0,1,\ldots,2B-2\}, \sum_{j=0}^M \rho_j \leq \sum_{j=0}^M \rho_{j}^* = \frac{2M+1}{4B}$. We know from~(\ref{B}) that $\sum_{j=0}^{K} j\rho_j \leq \sum_{j=0}^{2B} j\rho_j^*=B$. Also, since $\sum_{j=0}^{K} \rho_j = \sum_{j=0}^{2B} \rho_j^*=1$ and $\sum_{j=2B}^K \rho_j > \rho_{2B}^*$, it follows that there exists an integer $M \in \{0,1,\ldots,2B-1\}$ such that $\rho_M > \rho_M^*$; let $m$ be the smallest such integer. Since $\sum_{j=0}^m \rho_j \leq \sum_{j=0}^m \rho_j^*$, and $\forall j\in\{0,1,\ldots,m-1\}, \rho_j \leq \rho_j^*$, we can construct another message assignment by removing elements from some transmit sets whose size is $m$, such that the new assignment satisfies~(\ref{B}), and has transmit sets ${\cal T}_i^*$ where $\forall j\in\{0,1,\ldots,m\}, |\{i:i\in[K], |{\cal T}_i^*| = j\}| \leq \rho_j^*$. By successive application of the above argument, we can construct a message assignment that satisfies~(\ref{B}), and has transmit sets ${\cal T}_i^*$ where $\forall j\in\{0,1,\ldots,2B-1\}, |\{i:i\in[K], |{\cal T}_i^*| = j\}| \leq \rho_j^*$ and $|\{i:i\in[K], |{\cal T}_i^*| \geq 2B\}| \geq \rho_{2B}^*$. Note that the new assignment has to violate~(\ref{B}) since $\sum_{j=0}^{2B} j\rho_j^*=B$, and we reach a contradiction.

We now know from Lemma~\ref{lem:aux_ub} that under the backhaul load constraint of~(\ref{B}), the DoF for any $K$-user channel is upper bounded by $\frac{4B-1}{4B}K+o(K)$. It follows that the asymptotic per user DoF $\tau^{\text{avg}}(B) \leq \frac{4B-1}{4B}$, thereby proving the upper bound in Theorem~\ref{thm:main_result}.

\end{document}